\documentclass[aps,prd,nofootinbib,superscriptaddress,twocolumn]{revtex4-1}

\usepackage{amsmath,amssymb}
\usepackage{graphicx}
\usepackage{verbatim}
\usepackage{color}
\usepackage[colorlinks=true]{hyperref}
\usepackage{soul,xcolor}
\usepackage{multirow}

\setstcolor{red}

\hypersetup{linkcolor=black}

\newcommand{\be}{\begin{equation}}
\newcommand{\ee}{\end{equation}}
\newcommand{\ba}{\begin{eqnarray}}
\newcommand{\ea}{\end{eqnarray}}
\newcommand{\ban}{\begin{eqnarray*}}
\newcommand{\ean}{\end{eqnarray*}}

\newcommand{\nn}{\nonumber}

\numberwithin{equation}{section}

\begin{document}

\title{Bulk viscosity of strongly interacting matter  in the relaxation time approximation}

\author{Alina Czajka}
\affiliation{Department of Physics, McGill University, 3600 rue University,
Montreal, Quebec H3A 2T8, Canada}
\affiliation{Institute of Physics, Jan Kochanowski University,
Swietokrzyska 15 street, 
25-406 Kielce, Poland}

\author{Sigtryggur Hauksson}
\affiliation{Department of Physics, McGill University, 3600 rue
University,
Montreal, Quebec H3A 2T8, Canada}

\author{Chun Shen}
\affiliation{Department of Physics, Brookhaven National Laboratory, 
Upton, New York 11973-500, USA}

\author{Sangyong Jeon}
\affiliation{Department of Physics, McGill University, 3600 rue
University,
Montreal, Quebec H3A 2T8, Canada}

\author{Charles Gale}
\affiliation{Department of Physics, McGill University, 3600 rue
University,
Montreal, Quebec H3A 2T8, Canada}

\date{\today}

\begin{abstract}

We show how  thermal mean field effects
can be incorporated consistently in the hydrodynamical modeling of heavy-ion collisions. The nonequilibrium correction to 
the distribution function resulting from a temperature-dependent
mass is obtained in a procedure which automatically satisfies the Landau matching condition and is thermodynamically consistent. The physics of the bulk viscosity is studied here for  Boltzmann and Bose-Einstein gases
within the Chapman-Enskog and 14-moment approaches in the relaxation time
approximation. Constant and temperature-dependent masses are
considered in turn. It is shown that,  in the small mass limit, both methods lead to the same value of the ratio of the bulk
viscosity to its relaxation time. The
inclusion of a temperature-dependent mass leads to the emergence of the
$\beta_\lambda$ function in that ratio, and it is of the expected parametric form 
for the Boltzmann gas, while for the Bose-Einstein case it is affected by the infrared cutoff. This suggests that the relaxation time approximation may be  too crude to obtain a reliable form of $\zeta/\tau_R$ for gases obeying  Bose-Einstein statistics.

\end{abstract}


\maketitle

\section{Introduction}

The vibrant experimental programs pursued at the Relativistic Heavy
Ion Collider (RHIC) and at the 
Large Hadron Collider (LHC) have ushered in a new era of 
exploration of  systems governed by the nuclear strong interaction.  One of the remarkable features that emerged from investigating the physics of relativistic heavy-ion collisions is the fact that the created systems could be modeled theoretically by relativistic fluid dynamics  \cite{Gale:2013da,Heinz:2013th}. This realization led to developments in the formulation of relativistic viscous hydrodynamics in which observable consequences of the dissipative effects were isolated 
\cite{Baier:2007ix,Betz:2009zz,Romatschke:2009im,Kovtun:2012rj,Jeon:2015dfa,Florkowski:2017olj,Denicol:2011fa,Koide:2009sy,Denicol:2010xn,Denicol:2012es,Denicol:2012cn}. Currently, second-order viscous hydrodynamics
provides a description of the fluid behavior
\cite{Israel:1979wp,Denicol:2010xn,Denicol:2012es,Denicol:2012cn} which remedies the main failure of the Navier-Stokes -- or first-order -- formulation: acausal
signal propagation and numerical instabilities plaguing relativistic systems. 

While the 
hydrodynamic equations are universal and provide a macroscopic picture of a
relativistic fluid behavior in terms of conservation laws, transport
coefficients are governed by the underlying
microscopic theory which must be used for their extraction.
Although the first applications of viscous hydrodynamics focused on 
the shear viscosity, it has recently become clear that bulk viscosity
also plays an important role in the evolution of the QGP system
\cite{Ryu:2015vwa,Ryu:2017qzn,Paquet:2015lta}. The calculation of bulk viscosity from first principles,
however, remains a challenging  project. It is on this aspect that we concentrate in this paper.

The equations of the second-order hydrodynamics describe very efficiently
the expansion of the system produced in heavy-ion collisions.
This is a strong indication that the system must thermalize very
rapidly, which in turn indicates that the system is strongly interacting at
presently achievable energies.
Current estimates  of the bulk viscosity of QCD are mainly
based on the equation of state obtained from lattice QCD simulations \cite{Guenther:2017dou,Borsanyi:2016bzg}, or rely on empirical extractions based on simulations of relativistic nuclear collisions \cite{Ryu:2015vwa,Ryu:2017qzn,Paquet:2015lta,Paquet:2017mny}. Application of lattice QCD findings
\cite{Romatschke:2009ng,Karsch:2007jc,Hama:2005dz} and hadron resonance gas results
\cite{NoronhaHostler:2008ju,Denicol:2009am} made it possible to determine that the bulk viscosity
is notably enhanced near the critical temperature of the QCD phase
transition while the shear viscosity is substantially decreased in this
region \cite{Nakamura:2004sy,Csernai:2006zz}. 
Furthermore, the importance of bulk viscosity near
the transition temperature region was shown to have a remarkable impact on the elliptic
flow coefficient $v_2$ \cite{Song:2008hj,Denicol:2009am,Denicol:2010tr} and
other heavy-ion observables
\cite{Ryu:2015vwa,Ryu:2017qzn,Paquet:2015lta,Paquet:2017mny,Bozek:2017kxo,Monnai:2016kud}. Recently, the
behavior of bulk viscosity was also obtained from hydrokinetic theory,
which incorporates thermal noise \cite{Akamatsu:2017rdu}.

Despite the progress described above, there is still a need to develop methods which provide a better
insight in the effects of bulk viscosity at different energy
scales. In particular, one may be interested in having a consistent
analytical approach to bulk viscosity physics in the regime of very high
temperatures. At this energy scale the coupling constant is small and
fundamental quantum field theoretical tools can be used to study 
bulk viscosity systematically. Having
a comprehensive fluid dynamic formulation of a weakly coupled gas may also 
provide an essential benchmark for different approaches and
phenomenological applications.

In Refs.~\cite{Jeon:1994if,Jeon:1995zm} it was shown that quantum field
theory is equivalent, at least at leading order of perturbative
expansion, to kinetic theory. Later calculations then could 
use this efficient and intuitive kinetic theory
framework to study transport phenomena; see
\cite{Arnold:2000dr,Arnold:2002zm,Arnold:2003zc}. It has also provided a
natural language to formulate fluid dynamics concepts. Within the kinetic
approaches, the Chapman-Enskog and Grad's 14-moment methods are
commonly employed to study the nonequilibrium processes of a fluid. They,
however, rely on different treatments of the distribution function. While
the Chapman-Enskog theory deals directly with solving the Boltzmann
equation \cite{Chapman:1970}, Grad's approach is based on an expansion
of the nonequilibrium function in terms of the powers of momenta
\cite{Grad:1949}. To date, great progress has been made in extraction of
different transport coefficients within different theories. It seems,
however, that the comprehensive analysis of transport processes in a system
exhibiting conformal anomaly is not yet complete, especially in cases involving a mean field interaction.

A violation of conformal symmetry has a different impact on
different transport coefficients. It does not affect shear viscosity  much: its leading order behavior is dominated by the kinetic energy scale in weakly
interacting systems. On the other hand, the breaking of scale invariance 
dominates the physics of bulk viscosity. Consequently, the behavior of bulk
viscosity is largely determined by the sources of conformal symmetry breaking:
either the physical mass of plasma constituents or the Callan-Symanzyk 
$\beta_\lambda$ function, which fixes the coupling as a function of the
energy scale \cite{Jeon:1994if}. The parametric form of bulk viscosity
should then be dictated by the sources of scale invariance breaking squared,
as shown in Ref.~\cite{Arnold:2006fz} for QCD. The bulk viscosity of systems
exhibiting a conformal anomaly, due to the presence of a constant mass only, was
later studied within the Chapman-Enskog approach and the 14-moment
approximation, mostly in the relaxation time approximation \cite{Denicol:2014vaa,Jaiswal:2014isa,Florkowski:2015lra}, and
also within other approaches \cite{Huang:2010sa}. Moreover, quasiparticle
models were also examined for systems of various matter content in
Refs.~\cite{Sasaki:2008fg,Chakraborty:2010fr,Dusling:2011fd,Bluhm:2010qf,Romatschke:2011qp,Albright:2015fpa,Chakraborty:2016ttq,Tinti:2016bav,Alqahtani:2017jwl}. 

We observe, however, that there is still a need to revisit a
formulation of nonequilibrium fluid dynamics with the mean field background. 
Such a formulation is essential when one needs to include variable thermal masses 
consistently in the equations of viscous hydrodynamics. Having the correct form
of a nonequilibrium momentum distribution is also critical while studying 
some aspects of nuclear matter behavior phenomenologically, in particular,
when implementing the Cooper-Frye prescription in hydrodynamic simulations or 
examining electromagnetic probes in heavy-ion collisions 
\cite{Shen:2013cca,Shen:2014nfa,Paquet:2015lta,Hauksson:2016nnm}. 
Furthermore, such a consistent approach 
allows for an exhaustive calculation of transport coefficients.

The central part of this paper is devoted to derivation of the
nonequilibium correction to the distribution function where thermal effects
are consistently included. Subsequently, it is shown how the correction
influences the bulk viscosity behavior in the relaxation time
approximation. The analysis is done systematically and it comprises
different cases, namely, formulation of equilibrium and nonequilibrium
fluid dynamics and then computation of the ratio of  bulk viscosity to relaxation time. A computation is provided for gases of Boltzmann and Bose-Einstein statistics 
in both the Anderson-Witting model of the Chapman-Enskog method and the
14-moment approximation. The analysis performed in this paper is specific to  single-component bosonic degrees of freedom. Consequently, when the explicit forms of the thermal mass
and the $\beta_\lambda$ function are needed, we will use those of the 
scalar $\lambda\phi^4$ theory \cite{Jeon:1994if,Jeon:1995zm}. 
The method developed here is not appropriate for a one-component system 
following a Fermi-Dirac distribution function. Such a system would be a system of 
noninteracting fermionic degrees of freedom where the thermal mass and bulk viscosity 
cannot be determined. To count fermions accurately one needs to consider a
many-component system with the inclusion of  bosons mediating the interaction. This is not done here and is left for future work. 

The correction to the distribution function is found by noticing that there
is a twofold source of departure from equilibrium. First, there are hydrodynamic
forces that generate a deviation in the distribution function $\delta f$,
that is, they change the functional form of the distribution function. The
other source is related directly to interparticle interactions, the effect
of which is statistically averaged and emerges as the mean field.
Therefore, the correction is expressed by two terms; for the
Bose-Einstein gas the correction is
\ba
\Delta f = \delta f - T^2 \frac{d m^2_{\text{eq}}}{dT^2}
\frac{f_0(1+f_0)}{E_k}  
\frac{\int dK \delta f}{\int dK E_k f_0(1+f_0)}.\nn\\
\ea
For the description of quantities, see Table \ref{tab-quantities}. 
The obtained form of the correction allows one to formulate hydrodynamic
equations in a coherent way, where the Landau matching condition and
thermodynamic relations are guaranteed. Since the thermal mean field has a
negligible impact on shear viscosity, we further concentrate on bulk
viscosity dynamics, where the influence of the thermal background reveals
itself through the Landau condition and the speed of sound.

We show that both the Chapman-Enskog and 
the 14-moment approaches lead to the same final
expressions for the $\zeta/\tau_R$ ratio in the small mass limit, where $\tau_R$ is the bulk relaxation time. In general, 
temperature-dependent mass results in the emergence of the $\beta_\lambda$ function,
which dictates the very high temperature form of the ratio. In the
Boltzmann case the ratio is
\be
{\zeta_{\rm Boltz}\over \tau_R}
\approx
T^4 \left({1\over 3} - c_s^2\right)^2
\left(
{60\over \pi^2} - {36m_{x} \over \pi T}
\right),
\ee
where $(1/3-c_s^2)$ is directly related to $M_c$, the nonconformality parameter; see Table~\ref{tab-quantities}.
This shows the expected behavior of the source of scale invariance
breaking. 
One may observe that one factor of the scale invariance breaking
parameter is introduced directly by the Landau matching, which comes from
the small departure from equilibrium. The other factor emerges as a
correction to the pressure given by purely equilibrium quantities, but not
provided by the equation of state, as argued in \cite{Arnold:2006fz}. 
For a  system with Bose-Einstein statistics, the result is
\ba
{\zeta \over\tau_R}
\approx 
T^4 \left({1\over 3} - c_s^2\right)^2 
\left( {2\pi^3 T \over 25 m_{x}} - {4\pi^2 \over 75 } 
\left( 1-\frac{9 m^2_{\rm eq}}{8 m^2_x} \right) \right).\nn\\
\ea
The leading order term is not of the expected dependence because of the
factor $T/m_{x}$, which comes from an  infrared cutoff. The
same behavior is reflected if we neglect either the constant mass term or
thermally affected quantities. Therefore, it rather indicates that the relaxation time
approximation, which assumes that $\tau_R$ is energy independent, may not
allow one to entirely capture microscopic physics, in particular, of soft
momenta in quantum gases following a Bose-Einstein distribution function. 
A similar conclusion was reached in Ref.~\cite{Arnold:2006fz}.

The paper is organized as follows. In Sec.~\ref{sec-deviation} the
ingredients of the effective kinetic theory are briefly summarized and the
derivation of the noneqilibrium thermal correction is provided.
Section~\ref{sec-boltz-hydro} is devoted to the formulation of fluid dynamic basic
equations with the mean field background. In
Sec.~\ref{sec-bulk-ce}, the analysis of the ratio of bulk viscosity to relaxation time 
ratio is presented in the Chapman-Enskog theory, within which we solve the Anderson-Witting model. In Sec.~\ref{sec-evolution} we use the 14-moment approximation to derive the
evolution equation for the bulk pressure and then to calculate the bulk
viscosity over the relaxation time ratio and other transport coefficients in the bulk channel in the relaxation time approximation. 
Sec.~\ref{sec-summary} summarizes and concludes the work. Appendices
contain some technical details.

\begin{widetext}

\begin{table}[!h]
\centering
\begin{tabular}{lllll}
\hline \hline
Description &&  Equilibrium quantity && Nonequilibrium quantity 
\\
\hline \hline
Physical, zero-temperature mass of a particle && $m_0$ 
&& $m_0$ \vspace{1mm}
\\ 
Quasiparticle thermal mass 
&& 
$m_{\text{eq}}$ && $m_{\text{th}}$ \vspace{1mm}
\\
Quasiparticle mass && $m_x= \sqrt{m_0^2+ m_{\text{eq}}^2}$ 
&& 
$\tilde m_x= \sqrt{m_0^2+ m_{\text{th}}^2}$ \vspace{1mm}
\\
Quasiparticle energy && 
$E_k = \sqrt{{\bf k}^2+ m_x^2}$  \vspace{1mm}
&&
$\mathcal{E}_k = \sqrt{{\bf k}^2+ \tilde m_x^2}$  \vspace{1mm}
\\
Quasiparticle four-momentum && 
$k^\mu \equiv (k_0,{\bf k})=(E_k,{\bf k})$  
&&
$\tilde k^\mu \equiv (\tilde k_0,{\bf k})=(\mathcal{E}_k,{\bf k})$ \vspace{1mm}
\\
Lorentz invariant measure && 
$dK = d^3 {\bf k}/[(2\pi)^3 E_k]$ 
&&
$d\mathcal{K} = d^3 {\bf k}/[(2\pi)^3 \mathcal{E}_k]$ \vspace{1mm}
\\
Distribution function (in the local rest frame) && 
$f_0=1/[e^{\beta E_k}-1]$, with $\beta=1/T $
&& $f=f_0 + \Delta f$ \vspace{1mm}
\\
\hline \hline
Beta function for a coupling constant $\lambda$ & \multicolumn{4}{c}
{$\beta_\lambda = T d\lambda/dT=3\lambda^2/(16\pi^2)$ } 
  \vspace{1mm}
\\
Temperature dependence of the thermal mass & \multicolumn{4}{c}{$T^2 d m^2_{\text{eq}}/dT^2=m^2_{\text{eq}}+aT^2 \beta_\lambda$, with $a=1/48$ } 
\vspace{1mm}
\\
Nonconformality parameter &  \multicolumn{4}{c}{$M=(-m^2_0+aT^2 \beta_\lambda)/3$}
\\
\hline \hline
\end{tabular}
\caption{\label{tab-quantities} The quantities characterizing the
equilibrium and nonequilibrium dynamics of a gas with
Bose-Einstein statistics. For a classical gas with Boltzmann
statistics, some of these quantities have different values or forms, and whenever
there is a need to distinguish them we add the subscript $c$: $m_{\text{eq},c}$, $f_{0,c}=e^{-\beta E_k}$, $f_c$, $m_{\text{th},c}$, 
$a_c=1/(8\pi^2)$, and  $M_c$.}
\end{table}

\end{widetext}

\section{Nonequilibrium deviation from the equilibrium distribution
function}
\label{sec-deviation}

\subsection{Boltzmann equation with the mean field effect}
\label{sec-boltz}

Kinetic theory provides an efficient classical description of complex
microscopic dynamics of an interacting many-body system.
It is a good alternative to quantum field theory to study transport phenomena
 in the weakly coupled limit dominated by quasi-particle dynamics.
By quasiparticles one means particles which,
apart from zero temperature mass, gain additional thermal mass due to
interactions with the medium: the effect of the mean field.
They are characterized by a mean free path which is much larger than the
Compton wavelength of the system's constituents, and by a mean free time, which is
much larger than the time between collisions \cite{Arnold:2002zm}.
The dynamics of quasiparticles is encoded in the phase-space distribution
function which evolves according to the Boltzmann equation. 

We consider a system of uncharged thermally influenced particles of a
single species for which the Boltzmann equation reads
\ba
\label{boltz}
(\tilde k^\mu \partial_\mu -\mathcal{E}_k \nabla \mathcal{E}_k \cdot \nabla_k)
f=C[f],
\ea
where $C[f]$ is the collision term, $f=f(x,k)$ is a distribution function
of quasiparticles,\footnote{We use here such a notation that whenever $x$
and $k$ appear as arguments of a function, we mean $x^\mu$ and $\tilde k^\mu$ (or $k^\mu$ in the case of $f_0$),
respectively.} and the second term of the left-hand side involves the force
${\bf F}=d{\bf k}/dt = -\nabla \mathcal{E}_k$. The quasiparticle
four-momentum is defined as $\tilde k^\mu=(\tilde k^0, {\bf k})$, where $\tilde k_0 \equiv
\mathcal{E}_k$ is the nonequilibrium energy given by
\ba
\label{energy-noneq}
\mathcal{E}_k = \sqrt{{\bf k}^2+\tilde m_x^2},
\ea
which is a time- and space-dependent variable since $\tilde m_x^2 \equiv
\tilde m^2(x)=m_0^2+m^2_\text{th}(x)$, where $m_0$ is the physical
mass and $m_\text{th}(x)$ is the nonequilibrium thermal mass, which varies
in time and space. Knowing the $x$ dependence of the energy, one may rewrite
Eq.~(\ref{boltz}) as
\ba
\label{boltz-2}
\big(\tilde k^\mu \partial_\mu -\frac{1}{2} \nabla \tilde m_x^2 \cdot
\nabla_k\big) f=C[f].
\ea

The central object of the kinetic theory is the phase-space density
function $f(x,k)$. What we assume about the system is that its departure
from the equilibrium state is small, which, in turn, means that the process
of system equilibration is controlled by a small deviation in the
distribution function, which we denote as
\ba
\label{deltaf-0}
\Delta f(x,k) = f(x,k) - f_0(x,k),
\ea
where $f_0(x,k)$ is the equilibrium Bose-Einstein distribution function
and, in a general frame, it has the form
\ba
\label{distrib-function-zero-gen}
f_0(x,k) = \frac{1}{\exp[u_\mu(x)k^\mu(x)\beta(x)]-1},
\ea
where $\beta\equiv \beta(x)=1/T(x)$ with $T(x)$ being the local temperature, and
$u_\mu\equiv u_\mu(x)$ is the fluid four-velocity. The four-velocity in the
local rest frame is $u^\mu=(1,0,0,0)$. The quasiparticle four-momentum is
$k^\mu=(k^0, {\bf k})$, where $k^0$ component is the equilibrium
$x$-dependent energy 
\ba
E_k=\sqrt{{\bf k}^2+ m^2_x},
\ea 
where the dependence of $x$ enters through the mass $m^2_x \equiv m^2(x)=
m_0^2 + m^2_{\text{eq}}(x)$ with $m^2_{\text{eq}}(x)$ being the equilibrium
thermal mass, which is not the same as $m^2_{\text{th}}(x)$, the nonequilibrium thermal mass. The
Bose-Einstein density function in the fluid rest frame takes the form
\ba
\label{distrib-function-zero}
f_0(x,k) = \frac{1}{\exp\big( E_k(x)\beta(x) \big)-1}.
\ea
Let us add that in the forthcoming parts we will be deriving all equations
for the Bose-Einstein gas, but these equations may be analogously found for
the classical Boltzmann gas with the distribution function
\ba
\label{boltz-fun}
f_{0,c}(x,k)= \text{exp}(-\beta(x) u_\mu(x) k^\mu(x))
\ea 
and these will be briefly presented as well. Our aim is to reformulate the
equations of the viscous hydrodynamics when the effect of fluctuating
thermal mass is incorporated. Therefore, we assume that thermal influence
on the process of the system equilibration is controlled by the
nonequilibrium correction to the thermal mass, 
$\Delta m^2_{\text{th}} = m_{\rm th}^2 - m_{\rm eq}^2$,
which will be specified further. 

\subsection{Form of $\Delta f$}
\label{sec-ff}

As stated earlier, in this work we study systems with distribution functions that are perturbed from their equilibrium value. 
More specifically, the nonequilibrium phase space density can be written as 
\be
f(x,k) = f_{\rm th}(x,k) + \delta f(x,k)
\ee
The first part, $f_{\rm th}(x,k)$, still retains the local-equilibrium
form of the distribution function, but the thermal mass 
contains the nonequilibrium corrections
\ba
&&f_{\text{th}}(x,k) \equiv
\left. f_0(x,k)
\right|_{m_0^2 + m^2_{\text{eq}}(x)\to m_0^2 + m^2_{\text{eq}}(x)+\Delta m^2_{\text{th}}(x)}
\\
&& \;\;= \bigg[ \exp \Big( \sqrt{{\bf k}^2 + m^2_0 + m^2_{\text{eq}}(x) 
+ \Delta m^2_{\text{th}}(x)} \beta(x) \Big)-1\bigg]^{-1}.\nn
\ea
The second part, $\delta f(x,k)$,  is a change in the functional form of $f_0(x,k)$
caused by hydrodynamic forces, or equivalently,
nonvanishing gradients of energy and momentum densities.
The nonequilibrium correction $\Delta f$ then has two parts,
\ba
\label{Delta-f_not_used}
\Delta f(x,k) &=& f(x,k) - f_0(x,k) \nn \\
&=& \delta f(x,k) + \delta f_{\text{th}}(x,k),
\ea
where, to the leading order in small change,
$\delta f_{\rm th}(x,k) = f_{\rm th}(x,k) - f_0(x,k)$ is 
\ba
\label{delta-th}
\delta f_{\text{th}}(x,k) &=&
 -f_0(x,k) \big(1+f_0(x,k)\big) 
 \frac{\Delta m^2_{\text{th}}(x)}{2 E_k(x)} \beta(x), \qquad\;
\ea
which is obtained by expanding $f_{\rm th}$.
Since $\Delta m_{\rm th}^2$ is the nonequilibrium deviation,
it itself is going to be a functional of $\Delta f$.
Hence, the  equation
\ba
\label{Delta-f}
\Delta f = \delta f - \beta f_0(1+f_0) \frac{\Delta m_{\text{th}}^2}{2E_k}
\ea
must be solved self-consistently for $\Delta f$.

\subsection{Form of $\Delta m^2_{\rm{th}}$}
\label{sec-mm}

Recalling the basic foundations of effective kinetic theory, 
the analysis here relies
heavily on findings within the scalar $\lambda\phi^4$ theory,
as provided in Refs.~\cite{Jeon:1994if,Jeon:1995zm}, 
which makes the introduction of thermal
corrections analytically feasible. 
But the analysis presented here works equally well whenever the equilibrium
thermal mass has the form $\sim g^n T^2$, where $g$ is the dimensionless coupling
constant and $n$ is a positive integer.
We
intend to provide an effective macroscopic framework to study weakly
interacting systems, where the strength of interaction is determined by the
coupling constant $\lambda \ll 1$.
The coupling constant is scale
(temperature) dependent and the analysis performed here pertains only to
the perturbative regime. Within this approach the equilibrium thermal mass
is found to be
\ba
\label{mass-qq}
m^2_{\text{eq}} = \frac{\lambda(q_0)}{2} q_0,
\ea
where we have introduced the equilibrium scalar quantity $q_0$. The
function $q_0$ and its nonequilibrium counterpart $q$ are defined through
the corresponding distribution functions as
\ba
\label{q-fun-0}
q_0 &=& \int d K f_0,\\
\label{q-fun}
q &=& \int d \mathcal{K} f.
\ea
For the definitions of the symbols,
see Table\,\ref{tab-quantities}.
Therefore, one can observe that Eq.~(\ref{mass-qq}) contains the coupling
constant $\lambda(q_0)$, which is temperature dependent since $q_0$ is
temperature dependent.

Throughout the analysis we always keep the assumption that all
nonequilibrium quantities are slowly varying functions of space
points, which justifies that the nonequilibrium dynamics is governed by
small deviations of the quantities from their equilibrium values.
Therefore, we further assume that the nonequilibrium thermal mass is a
function of the scalar quantity $q$ only. 
The same assumption is applied to the running coupling $\lambda(q)$.
Thus, the nonequilibrium thermal mass can be
expanded as
\ba
\label{mas-del}
m^2_{\text{th}}(q) = m^2_{\text{th}}(q_0+\Delta q) 
= m^2_{\text{eq}}(q_0) + \Delta m_{\text{th}}^2
\ea
with
\ba
\label{mas-del-q}
\Delta m_{\text{th}}^2 = \frac{d m^2_{\text{eq}}}{dq_0} \Delta q.
\ea
The function $q$ is uniquely defined by Eq.~(\ref{q-fun}) and should be
obtained self-consistently from this 
equation. Hence to evaluate $\Delta m_{\text{th}}^2$, we need to find
$\Delta q$ which is itself a function of $ \Delta
m_{\text{th}}^2$. The 
deviation of the scalar quantity $q$ can be written as 
\ba
\Delta q &=&\int dK \delta f +  \frac{\partial q_0}{\partial
m^2_{\text{eq}}} \Delta m_{\text{th}}^2.
\ea
Equation~(\ref{mas-del-q}) then takes the form
\ba
\label{mas-del1}
\Delta m^2_{\text{th}} = \frac{1}{1- \frac{dm^2_{\text{eq}}}{dq_0}
\frac{\partial q_0}{\partial m^2_{\text{eq}} }}
\frac{dm^2_{\text{eq}}}{dq_0} \int dK \delta f.
\ea
On the other hand both $m^2_{\text{eq}}$ and $q_0$ are related by
temperature, so that one can find
\ba
&&\frac{dm^2_{\text{eq}}}{dT}= \frac{dm^2_{\text{eq}}}{dq_0} \frac{dq_0}{dT} \nn\\
&& \;\;=\frac{dm^2_{\text{eq}}}{dq_0} \bigg( \beta^2 \int dK E_k f_0(1+f_0) +
\frac{dm^2_{\text{eq}}}{dT} \frac{\partial q_0}{ \partial m^2_{\text{eq}}}
\bigg).\qquad
\ea
Extracting further $\frac{dm^2_{\text{eq}}}{dq_0} \frac{\partial
q_0}{\partial m^2_{\text{eq}} }$ and inserting it to Eq.~(\ref{mas-del1})
leads to 
\ba
\label{Delta-m}
\Delta m^2_{\text{th}} =2 T^2 \frac{d m^2_{\text{eq}}}{dT^2} \frac{\int dK
\delta f}{\beta \int dK E_k f_0(1+f_0)},
\ea
where we used $d m^2_{\text{eq}}/dT = 2T d
m^2_{\text{eq}}/ dT^2$.
\vspace{0.2cm}

Inserting Eq.~(\ref{Delta-m}) into
Eq.~(\ref{Delta-f}), one gets 
\ba
\label{Delta-f1}
\Delta f = \delta f - T^2 \frac{d m^2_{\text{eq}}}{dT^2}
\frac{f_0(1+f_0)}{E_k}  
\frac{\int dK \delta f}{\int dK E_k f_0(1+f_0)}.\qquad\;
\ea
Analogously, the correction for the Boltzmann gas is
\ba
\label{Delta-f1-bol}
\Delta f_c = \delta f_c - T^2 \frac{d m^2_{\text{eq},c}}{dT^2}
\frac{f_{0,c}}{E_k}  
\frac{\int dK \delta f_c}{\int dK E_k f_{0,c}},
\ea
where the subscript $c$ has been used to emphasize that the formula holds
for the classical gas.
Equations~(\ref{Delta-f1}) and (\ref{Delta-f1-bol}) are one of the main results of 
this paper. In previous analyses \cite{Sasaki:2008fg,Chakraborty:2010fr,Bluhm:2010qf,Romatschke:2011qp,Albright:2015fpa,Chakraborty:2016ttq,Tinti:2016bav,Alqahtani:2017jwl}, the second term in Eq.~(\ref{Delta-f1}) was missing or was incomplete. When applying the Cooper-Frye formula in viscous hydrodynamics,
it is $\Delta f$, not $\delta f$ that should be used.

\subsection{Temperature dependence of the thermal mass}
\label{sec-tm}

The thermal mass is a function of the
scalar quantity $q_0$ and is defined by Eq.~(\ref{mass-qq}). 
Its temperature dependence is dictated by
\ba
\label{mass-t}
\frac{dm^2_{\text{eq}}}{dT} = \frac{\lambda(q_0)}{2} \frac{dq_0}{dT}
+ \frac{q_0}{2} \frac{d\lambda(q_0)}{dT}.
\ea
$q_0$ is one of the thermodynamic functions discussed in detail in Appendix
\ref{bessel-fun}, and its leading order value is found to be $T^2/12$.
Additionally, the second term in Eq.~(\ref{mass-t})
encodes the running of the coupling constant as a function of the energy
scale, which is the essence of the renormalization group
$\beta_\lambda$ function, defined by
\ba
\beta_\lambda \equiv \beta(\lambda) = T \frac{d\lambda(q_0)}{dT}.
\ea
It should be obtained using diagrammatic methods. 
In the case of scalar theory,
$\beta_\lambda$ is positive and proportional to $\lambda^2$. 
Collecting these contributions, one finds
\ba
\label{mass-temp}
T^2\frac{dm^2_{\text{eq}}}{dT^2} = m^2_{\text{eq}} + a T^2 \beta_\lambda.
\ea
where $m^2_{\text{eq}} = \lambda T^2/24$ and $a=1/48$.

One can analogously consider a temperature-dependent scaling for the
classical Boltzmann gas. In this case, the thermal effective mass may 
be assumed to have the same form as (\ref{mass-qq}). 
The only difference is that one uses the Boltzmann distribution function
$f_{0,c}$ instead of $f_0$. This gives
$q_{0c} = T^2/(2\pi^2) + O(m_x^2)$, as given
by Eq.~(\ref{I00}), and it leads to
\ba
\label{mass-temp_2}
T^2\frac{dm^2_{\text{eq},c}}{dT^2} = m^2_{\text{eq},c} + a_c T^2
\beta_\lambda,
\ea
where $m^2_{\text{eq},c} = \lambda T^2/(4\pi^2)$ and $a_c=1/(8\pi^2)$.

\section{Equations of hydrodynamics with thermal corrections}
\label{sec-boltz-hydro}

\subsection{Local equilibrium hydrodynamics}
\label{sec-hydro-eq}

First consider a system under strict local equilibrium.
By that we mean that the functional form of the distribution function
is still $f_0$ given in Eq.~(\ref{distrib-function-zero-gen})
or in Eq.~(\ref{boltz-fun}),
but the temperature as well as the thermal mass are $x$ dependent.
Such a system 
possesses a conserved stress-energy tensor of the form
\ba
\label{T-zero}
T_0^{\mu\nu} = \int dK k^\mu k^\nu f_0 -g^{\mu\nu}U_0,
\ea
where the metric tensor we use is $g^{\mu\nu} = \text{diag}(1,-1,-1,-1)$.
The extra term $U_0\equiv U_0(x)$ is the mean-field contribution that guarantees the
thermodynamic consistency of hydrodynamic equations and the conservation of
energy and momentum, via the following condition: 
\ba
\label{U-zero}
dU_0 = \frac{q_0}{2} dm^2_{\text{eq}},
\ea
where $q_0$ is the Lorentz scalar defined by Eq.~(\ref{q-fun-0}). 

Since we study here a system with no conserved charges, the
Landau frame is a natural kinetic framework to define the four-velocity
$u^\mu$ via
\ba
\label{eigen}
u_\mu T_0^{\mu\nu} = \epsilon_0 u^\nu,
\ea
where the eigenvalue $\epsilon_0$ can be identified as the
local energy density. With this definition the energy-momentum tensor may
be decomposed using two orthogonal projections $u^\mu u^\nu$ and
$\Delta^{\mu\nu} = g^{\mu\nu} - u^\mu u^\nu$.
The equilibrium energy-momentum tensor becomes
\ba
\label{tensor-zero}
T_0^{\mu\nu} = \epsilon_0 u^\mu u^\nu - P_0 \Delta^{\mu\nu},
\ea
where $P_0$ is the local thermodynamic pressure. The energy density and the
pressure are in turn given by
\ba
\label{energy-pressure}
\epsilon_0 &=& \bar \epsilon_0- U_0, \\
\label{energy-pressure2}
P_0 &=& \bar P_0+U_0,
\ea
where 
\ba
\bar \epsilon_0 &=& \big \langle (u_\mu k^\mu)^2 \big \rangle_0, \\
\bar P_0 &=& -\frac{1}{3} \big \langle \Delta^{\mu\nu} k_\mu k_\nu \big \rangle_0
\ea
with the notation $\big \langle \dots \big \rangle_0 = \int dK \dots f_0$.
Let us point out that the enthalpy is not changed by the mean field 
$\bar\epsilon_0+\bar P_0 = \epsilon_0+ P_0 $. One may also check that the
definitions of energy density (\ref{energy-pressure}) and pressure (\ref{energy-pressure2}), together
with the condition (\ref{U-zero}), guarantee that the thermodynamic
relation
\ba
T s_0=T\frac{d P_0}{dT} = \epsilon_0+ P_0,
\ea
where $ s_0$ is the entropy density, is satisfied.

\subsection{Nonequilibrium hydrodynamics}
\label{sec-hydro-non}

The stress-energy tensor of fluid dynamics out of equilibrium takes the
following form:
\ba
\label{T-noneq}
T^{\mu\nu} = \int d\mathcal{K} \tilde k^\mu \tilde k^\nu f -g^{\mu\nu}U,
\ea
which is formally the same as Eq.~(\ref{T-zero}).
The mean-field correction $U$ must be now a function of 
$q = \int d{\cal K} f$ only \cite{Jeon:1995zm}.
We emphasize that the formulation of the fluid
hydrodynamic framework with the thermal correction still has to
conform with all assumptions that were made to provide the effective kinetic
theory, discussed in Sec. \ref{sec-deviation}.
In particular, such a
description requires the system to be sufficiently dilute and the
quasiparticles' mean free paths to be much longer than the thermal width of
its constituents, which is maintained when the strength of interaction is
weak. Furthermore, to allow for validity of hydrodynamics, the system has
to be characterized by some macroscopic length scale at which macroscopic
variables, such as pressure and energy density, vary. Under these
assumptions, a nonequilibrium hydrodynamic description applies to systems
where departures of all quantities from their equilibrium values are
characterized by small corrections. Therefore, the nonequilibrium function
$U$, in particular, may be expanded as
\ba
\label{U-Del}
U=U_0 + \Delta U ,
\ea
where
\ba
\label{U-Delq}
\Delta U = \frac{dU_0}{dq_0} \Delta q. 
\ea
However, as discussed before and explicitly 
shown by Eqs.~(\ref{mas-del}) and (\ref{mas-del-q}), the thermal mass is
also a function of $q$ only. Therefore, applying the relation
(\ref{mas-del-q}) to (\ref{U-Delq}), one finds
\ba
\label{U1}
\Delta U 
= \frac{q_0}{2}  \Delta m^2_{\text{th}}.
\ea
As before, this is also the condition that $U$
must satisfy to maintain the energy-momentum conservation law $\partial_\mu
T^{\mu\nu}=0$.

The stress-energy tensor of the viscous hydrodynamics (\ref{T-noneq}) may
be next decomposed into the local equilibrium part and the nonequilibrium
deviation 
\ba
\label{tensor-delta}
T^{\mu\nu} = T^{\mu\nu}_0 + \Delta T^{\mu\nu},
\ea
where $T^{\mu\nu}_0$ is given by (\ref{tensor-zero}) and $\Delta
T^{\mu\nu}$ carries all dynamical
information needed in order to determine how the nonequilibrium system
evolves into equilibrium. Note that a separation of the viscous
correction from the equilibrium part in Eq.~(\ref{tensor-delta}) has
been done not as a rearrangement of Eq.~(\ref{T-noneq}) but rather as an
expansion of the stress-energy tensor around its local equilibrium value. 
As shown in Appendix \ref{components}, we have
\ba
\label{T-00}
\Delta T^{00} &=& \int dK E_k^2 \Delta f, \\
\label{T-0i}
\Delta T^{0i} &=& \int dK E_k k^i \Delta f , \\
\label{T-ij}
\Delta T^{ij} &=& 
\int dK k^i k^j \Delta f 
-\frac{\Delta m^2_{\text{th}}}{2} \int dK \frac{k^i k^j}{E_k^2} f_0 \nn\\
&&+\delta^{ij} \frac{\Delta m^2_{\text{th}}}{2}
\int dK f_0,
\ea
where $\Delta m^2_{\text{th}}$ and $\Delta f$ are given by (\ref{Delta-m})
and (\ref{Delta-f1}), respectively. Equations (\ref{T-00}) and (\ref{T-0i})
shall dictate the form of the Landau matching condition,
and Eq.~(\ref{T-ij}) contains the definitions of the viscous corrections.

\subsection{Landau matching condition in the rest frame}
\label{sec-hydro-landau}

The Landau matching is defined by the eigenvalue problem
\ba
\label{eigen-non}
u_\mu T^{\mu\nu} = \epsilon u^\nu,
\ea
where $\epsilon$ is the energy density of the nonequilibrium state
including the thermal correction $U$. In the
fluid rest frame it comes down to two equations, corresponding to the
conditions on the energy density and the momentum density:
\ba
T^{00} = \epsilon, \qquad\qquad T^{0i} = 0.
\ea
Under the Landau matching condition, 
the local equilibrium is defined to have the same local energy and the momentum density
\ba
\label{T-00-landau}
\Delta T^{00} = 0,
\qquad\qquad\qquad
\label{T-0i-landau}
\Delta T^{0i} = 0.
\ea
Using Eqs.~(\ref{T-00}) and (\ref{T-0i}) with the correction to the
distribution function $\Delta f$
given by Eq.~(\ref{Delta-f1}), we obtain
\ba
\label{landau-1}
\Delta \epsilon \! &=&\! \int \! dK \bigg[E_k^2 - T^2
\frac{dm^2_{\text{eq}}}{dT^2} \bigg] \delta f , \\
\label{landau-2}
0 \! &=& \! \int \! dK \bigg[E_k k^i - T^2 \frac{dm^2_{\text{eq}}}{dT^2} 
 \! \frac{\int \!dK' k^{\prime i} f_0(f_0+1)}{\int \!dK' E_k' f_0(f_0+1)}\bigg] \delta f.\qquad\;
\ea
However, the second term in Eq.~(\ref{landau-2}) vanishes because of rotational symmetry in 
equilibrium. 
Hence the Landau matching conditions are
\ba
\label{landau-e}
\int dK \bigg[E_k^2 - T^2 \frac{dm^2_{\text{eq}}}{dT^2} \bigg] \delta f &=&
0 , \\
\label{landau-p}
\int dK E_k k^i  \delta f &=& 0.
\ea
The second condition indicates that $\delta f$ cannot have
a vector component: it can only contain a spin 0 part
and a spin 2 part.

\subsection{Shear-stress tensor and bulk pressure in the local rest frame}
\label{sec-hydro-viscous}

The shear tensor $\pi^{ij}$ and the bulk pressure $\Pi$ are found from Eq.~(\ref{T-ij}) in the local rest frame, where
 Eqs.~(\ref{Delta-m}) and (\ref{Delta-f1}) are inserted. Then, as shown
in Appendix~\ref{components}, one obtains
\ba
\label{T-ij-del}
\Delta T^{ij} = \int dK k^i k^j \delta f.
\ea
We can reorganize (\ref{T-ij-del}) to separate
the spin 0 part and the spin 2 part as follows:
\ba
\Delta T^{ij} = \pi^{ij} + \delta^{ij} \Pi,
\ea
where
\ba
\label{pi-ij}
\pi^{ij} &=& \int dK k^{\langle i} k^{j \rangle} 
\delta f, \\
\label{Pi-expr}
\Pi &=& \frac{1}{3} \int dK {\bf k}^2 \delta f,
\ea
where $k^{\langle i}k^{j\rangle} = k^i k^j - {\bf k}^2\delta^{ij}/3$.
These coincide with the commonly known forms of the shear-stress tensor and bulk
pressure in the local rest frame.

\subsection{General frame}
\label{sec-hydro-gen}

In a general frame where the flow velocity $u^\mu$ may be arbitrary, 
the energy-momentum tensor 
is\footnote{In Ref.~\cite{Jeon:1994if}, the energy-momentum tensor
correction was written down incorrectly, but the mistake vanished with the imposition of the Landau matching condition, ensuring the validity of the subsequent
derivations.}
\ba
T^{\mu\nu} &=& \int dK k^\mu k^\nu f_0 - g^{\mu\nu}U_0 \nn\\
&&
+ \int dK \bigg[ k^\mu k^\nu -u^\mu u^\nu T^2\frac{dm^2_{\text{eq}}}{dT^2}
\bigg] \delta f.
\ea
The Landau condition then becomes
\ba
 \int dK \bigg[ (u_\mu k^\mu) k^\nu - u^\nu
T^2\frac{dm^2_{\text{eq}}}{dT^2} \bigg]\delta f =0
\ea
and the viscous corrections are given by
\ba
\label{viscous-corr}
\pi^{\mu\nu} = \big\langle  k^{\langle \mu} k^{\nu \rangle}
\big\rangle_\delta ,
\qquad
\label{viscous-pi}
\Pi  =  -\frac{1}{3}\big\langle  \Delta_{\mu\nu} k^\mu k^\nu
\big\rangle_\delta,
\ea
where $\langle \dots \rangle_\delta \equiv \int dK (\dots) \delta f$. We
have also used the notation 
$A^{\langle \mu\nu \rangle} \equiv \Delta^{\mu\nu}_{\alpha \beta}
A^{\alpha\beta}$, where 
$\Delta^{\mu\nu}_{\alpha \beta} \equiv (\Delta^\mu_\alpha \Delta^\nu_\beta
+ \Delta^\mu_\beta \Delta^\nu_\alpha - 2/3
\Delta^{\mu\nu}\Delta_{\alpha\beta})/2$.
The definitions (\ref{viscous-corr}) have
well-known structures, but the thermal mass that enters
them is now $x$ dependent and the Landau matching contains a correction due
to the temperature-dependent mass. These arguments are essential when one
aims at examining transport properties of the medium.

\section{Nonequilibrium correction in the Chapman-Enskog approach}
\label{sec-bulk-ce}

Chapman-Enskog theory provides 
a way to directly find the solution to the
Boltzmann equation for near-equilibrium systems. 
Solving the full Boltzmann equation, however, is formidable task.
In this paper, we use the Anderson-Witting model \cite{Anderson:1973} to find
the explicit leading order solution.
In this section, we focus on the bosonic quantum gas case.
Treatment for the Boltzmann gas case is identical
if one replaces $f_0(1+f_0)$ with the Boltzmann factor $f_{0,c}$.

\subsection{Solution of the Anderson-Witting equation in the rest frame}

With the medium-dependent thermal mass, the Anderson-Witting model is
given by
\be
\left( \tilde k^\mu\partial_\mu - {\cal E}_k \nabla {\cal E}_k\cdot \nabla_k \right)f = 
-{(u\cdot \tilde k)\over \tau_R}\Delta f,
\label{eq:AW_eq}
\ee
where $\tilde k^\mu = ({\cal E}_k, {\bf k})$. In the fluid cell rest frame 
$u^\mu = (1, 0, 0, 0)$ and $u\cdot \tilde k = {\cal E}_k$.

To use the Chapman-Enskog method, we let
\be
f = f_0 + f_1 + f_2 + \cdots
\ee
where each $f_n$ contains only the $n$-th derivatives of
the thermodynamic quantities and the flow velocity.
The first-order equation is obtained by identifying $\Delta f = f_1$
 in the right-hand side and using all other quantities in their equilibrium forms
\be
\left(k^\mu\partial_\mu 
- {1\over 2}\partial_i m_{\rm eq}^2 
{\partial\over \partial k_i}\right)f_0(x,k)
= -{E_k\over \tau_R} \Delta f(x,k),
\ee
where now $k^\mu = (E_k, {\bf k})$.

Evaluating the left-hand side yields
\ba
&&
\Big( k^\mu \partial_\mu - {1\over 2}\partial_i m_{\rm eq}^2 
{\partial \over \partial k_i}\Big)f_0(x,k)
=
-\beta 
f_0(x,k)(1 + f_0(x,k)) \nn \\
&& \;\;\;\times \bigg[
\bigg(
c_s^2\left(E_k^2 - T^2{dm_{\rm eq}^2\over dT^2}\right)
- {{\bf k}^2\over 3}
\bigg)
(\partial_i u^i)
- 
k^{\langle j} k^{i\rangle} \partial_j  u_i 
\bigg], \nn \\
\label{eq:AW-LHS}
\ea
where the equations of motion from the ideal hydrodynamics
\ba
\label{part-u}
\partial_0 u^i  &=& \frac{\partial^i T}{T},\\
\label{dtemp}
\partial_0 T &=& - T c_s^2 \partial_i u^i
\ea
are used to remove time derivatives.

The $\Delta f$ in the right-hand side of the Anderson-Witting model
is just Eq.~(\ref{Delta-f1}).
Letting $\delta f = f_0(1+f_0)\phi$, we get
\ba
&&
\Delta f(k)
=
f_0(k)(1 +f_0(k)) \nn \\
&&\times \left(\phi(k) - 
{T^2\over E_k} {dm_{\rm eq}^2\over dT^2} 
{\int dK\phi(k) f_0(k)(1 + f_0(k)) 
\over \int dK E_{k}f_0(k)(1 + f_0(k))}
\right)\!\!,\;\;\;\;\;
\label{eq:AW-RHS}
\ea
where the $x$ dependence of all quantities is suppressed for the sake of brevity.
In previous derivations, the last term 
was missing \cite{Bluhm:2010qf,Tinti:2016bav,Alqahtani:2017jwl}. 
Dividing $\phi$ into the shear and the bulk parts 
$\phi = \phi_{\rm s} + \phi_{\rm b}$, and comparing
Eqs.~(\ref{eq:AW-LHS}) and (\ref{eq:AW-RHS}),
the shear part of $\phi$ is trivially obtained as
\be
\phi_{\rm s}(k)
= 
-{\tau_R \over TE_k}
k^{\langle j} k^{i\rangle} \partial_j  u_i ,
\label{phi_s}
\ee
since the angle integration over the spin-2 tensor $k^{\langle j}k^{i\rangle}$ 
vanishes.
For the bulk part, 
letting
\be
\phi_{\rm b}(k) = \left(aE_k + {b\over E_k}\right) \partial_i u^i
\ee
and comparing Eqs.~(\ref{eq:AW-RHS}) and (\ref{eq:AW-LHS}),
we get
\be
a = \tau_R \beta \left(c_s^2 - {1\over 3}\right)
\label{a-expr}
\ee
and
\ba
b
&=&
{-M\beta\tau_R J_{1,0} \over J_{1,0} - T^2(dm_{\rm eq}^2/dT^2) J_{-1,0}},
\label{b-expr}
\ea
where we defined
\be
M = -{1\over 3}\left(m_x^2 - T^2{dm_{\rm eq}^2\over dT^2}\right) .
\label{eq:M}
\ee
With $m_{\rm eq}^2 \propto \lambda T^2$, we have 
\be
M = -{1\over 3}\left(m_0^2 - a\beta_\lambda T^2\right),
\ee
where $\beta_\lambda$ is the coefficient function of the coupling constant
renormalization group and $a = O(1)$ depends on the theory.
The parameter $M$ 
can be identified as the parameter of nonconformality of the system (or
the source of the conformal invariance violation). 
We have also introduced a notation for thermodynamic integrals,
\ba
\label{J}
J_{n,q} &=& a_q\int dK  (u \cdot k)^{n-2q} 
(-\Delta_{\mu\nu} k^\mu k^\nu )^q \, f_0(k)(1+f_0(k)),\nn\\
&&
\ea
where $a_q=1/(2q+1)!!$, which can be evaluated in the fluid cell rest frame.
The bulk part of the leading order Chapman-Enskog solution
of the Anderson-Witting equation is then
\ba
\label{cs2}
&&\phi_{\rm b}(k)=\tau_R \beta (\partial_i u^i) \nn\\
&&\times \left(
(c_s^2 - 1/3)E_k 
-{1\over E_k}
{M J_{1,0}\over J_{1,0} - T^2(dm_{\rm eq}^2/dT^2) J_{-1,0}}
\right).\nn\\
&&
\ea
To show that $\phi_{\rm b}(k)$ is in fact proportional to $(c_s^2-1/3)$,
we can use
\ba
\label{sound}
c_s^2 = \frac{d P_0/dT}{d \epsilon_0/dT}
= \frac{J_{3,1} }
{J_{3,0} - (T^2 d{m_{\rm eq}^2/dT^2})J_{1,0}},
\ea
where $ P_0$ and $ \epsilon_0$ 
are the pressure and the energy density
given in Eqs.~(\ref{energy-pressure2}) and (\ref{energy-pressure}).
Using the identities from Appendix~\ref{bose-details}, one can also show
that
\ba
{1\over 3} - c_s^2
&=&
-
\frac{ MJ_{1,0} }
{ 
J_{3,0} - T^2(dm_{\rm eq}^2/dT^2)J_{1,0 } 
}.
\label{speed-be}
\ea
Hence finally
\ba
\label{eq:phi_b}
&&\phi_{\rm b}(k) =
\tau_R \beta(\partial_i u^i)(c_s^2 - 1/3) \nn\\
&&\qquad \times \left(E_k -{1\over E_k}{J_{3,0} - T^2(dm_{\rm eq}^2/dT^2)J_{1,0}\over J_{1,0} 
- T^2(dm_{\rm eq}^2/dT^2) J_{-1,0}}
\right).\qquad
\ea
Equation~(\ref{eq:phi_b}) is another main result in this work. This equation slightly differs from the analogous one for the Boltzmann statistics shown in Ref.~\cite{Romatschke:2011qp,Chakraborty:2010fr,Paquet:2015lta}.

In hydrodynamic simulations, it is practical to replace the system expansion rate by the bulk viscous pressure using the Navier-Stokes relation $\Pi = - \zeta \theta$, which gives
\ba
&&\phi_{\rm b}(k) =
\beta \left( - \frac{\Pi}{\zeta/\tau_R} \right)(c_s^2 - 1/3) \nn\\
&&\qquad \times \left(E_k -{1\over E_k}{J_{3,0} - T^2(dm_{\rm eq}^2/dT^2)J_{1,0}\over J_{1,0} 
- T^2(dm_{\rm eq}^2/dT^2) J_{-1,0}}
\right).\qquad
\label{eq:phi_b_fin}
\ea

Having given the solution of the Anderson-Witting equation, one can also find $\Delta f$ explicitly. Inserting Eqs.~(\ref{eq:phi_b}) and (\ref{phi_s}) into (\ref{eq:AW-RHS}) one finds
\ba
\label{del-II}
\Delta f(k)&=&f_0(k)(1+f_0(k))\tau_R \beta
\bigg[-(\partial_j  u_i) \frac{ k^{\langle j} k^{i\rangle} }{E_k}  \nn\\
&&+ (\partial_i u^i)(c_s^2-1/3)
\bigg(E_k - \frac{1}{E_k}\frac{J_{3,0}}{J_{1,0}}\bigg)\bigg].\;\;
\ea
The phase space density correction $\Delta f$ has a much
simpler form than $\phi$. However, for transport coefficient calculations, it is
$\phi$ (equivalently $\delta f$),
rather than $\Delta f$, that is needed.

\subsection{Energy conservation and Landau matching in the Anderson-Witting case}

By multiplying $\tilde k^\nu = ({\cal E}_k, {\bf k})$ and integrating over $d{\cal K}$, 
the left-hand side of Anderson-Witting equation (\ref{eq:AW_eq}) turns into
$\partial_\mu T^{\mu\nu}$, where the stress-energy tensor $T^{\mu\nu}$ is defined in Eq.~(\ref{T-noneq}). Assuming that the mean-field contribution $U$ satisfies
\be
\partial_\mu U(x) = {\partial_\mu \tilde{m}_{x}^2(x)\over 2}\int d{\cal K} f(x,k),
\ee
we get $\partial_\mu T^{\mu\nu} = 0$.

Under the same condition, the right-hand side of the 
Anderson-Witting model within the Chapman-Enskog approach must also vanish,
\be
-{1\over \tau_R} \int dK\, E_k k^\mu\, \Delta f = 0,
\ee
to ensure energy-momentum conservation.
This condition for energy-momentum conservation is actually
exactly the same as the Landau conditions we derived in 
Sec.~\ref{sec-hydro-landau}.
Upon using $\Delta f$ in Eq.~(\ref{Delta-f1}) in the fluid rest frame,
these become
\be
0 = \int dK \left(E^2_k - T^2{dm_{\rm eq}^2\over dT^2} \right)\delta f
\label{eq:Landau_cond_ene}
\ee
and
\be
0 
=
\int  dK E_k k^i \delta f .
\label{eq:Landau_cond_pi}
\ee
Eq.~(\ref{eq:Landau_cond_pi}) is automatically satisfied
by the $\delta f = f_0(1+f_0)(\phi_{\rm s} + \phi_{\rm b})$ obtained
in the previous subsection since it does not contain a vector part.
In the condition (\ref{eq:Landau_cond_ene}), the shear part $\phi_{\rm s}$
also vanishes because it contains a spin-2 tensor.
Using Eqs.~(\ref{eq:phi_b}) and (\ref{J}), it is easy to show
that the energy conservation and the Landau condition are indeed fulfilled.
This automatic fulfillment of the Landau condition for the quasiparticle case 
would not have been possible if one missed the $\Delta m^2_{\rm th}$ correction in $\Delta f$.

\subsection{The shear and the bulk viscosities in the Anderson-Witting model}
\label{sec-ce-ratio}

The full leading order Chapman-Enskog solution to the Anderson-Witting model is
given by Eq.~(\ref{eq:AW-RHS}) with $\phi_{\rm s}$ and $\phi_{\rm b}$ obtained above.
The shear viscosity can be evaluated by using Eq.~(\ref{pi-ij}) for $\pi^{ij}$
and Eq.~(\ref{phi_s}) for $\phi_{\rm s}$ as
\be
\pi^{ij} = 
\frac{2\beta}{15} \tau_R \int dK\, f_0(1+f_0)\frac{{\bf k}^4}{E_k} \sigma^{ij},
\ee
where $\sigma^{ij}=-1/2(\partial^i u^j + \partial^j u^i -2/3 g^{ij} \partial_k u^k)$.
Identifying $\pi^{ij} = 2\eta \sigma^{ij}$,
we get
\be
{\eta\over\tau_R} = \beta J_{3,2}
\ee
and subsequently find the shear viscosity in the relaxation time
approximation, which was examined in few papers, see, for example,
\cite{Romatschke:2011qp,Denicol:2014vaa,Florkowski:2015lra}, and has the form
\ba
\label{shear-v}
\frac{\eta}{\tau_R} = \frac{\epsilon_0+P_0}{5}.
\ea

For the bulk viscosity, we start with Eq.~(\ref{Pi-expr})
\be
\Pi = \int dK\, {{\bf k}^2\over 3}\, \delta f.
\ee
Using the Landau condition, Eq.~(\ref{landau-e}), one gets
\ba
\Pi &=& M \int dK \delta f,
\label{eq:Pi_final}
\ea
in which only the bulk part is relevant:
\ba
\Pi 
&=&
M\int dK\, f_0(k) (1+f_0(k))\phi_b(k)
\label{eq:Pi_bulk}
\ea
with $\phi_b(k)$ given by Eq.~(\ref{eq:phi_b}). Since $\Pi = -\zeta \partial_i u^i$,
one can read off the ratio of bulk viscosity to the relaxation time from Eq.~(\ref{eq:Pi_bulk}) as
\ba
&&{\zeta\over \tau_R}
=
\beta M^2 
\bigg(
{J_{-1,0}J_{1,0}\over J_{1,0} - T^2(dm_{\rm eq}^2/dT^2) J_{-1,0}} \qquad\qquad\nn \\
&&\qquad\qquad\qquad
-\frac{ J_{1,0} J_{1,0} } {J_{3,0} - T^2(dm_{\rm eq}^2/dT^2)J_{1,0}}
\bigg).
\label{bulk-be1}
\ea
The integrals present in Eq.~(\ref{bulk-be1}) have been computed 
in Appendix \ref{bose-details}. Using them, one gets the value of the ratio as
\ba
\label{bulk-be-2-ce}
\frac{\zeta}{\tau_R} \approx 
\frac{M^2}{2\pi^2}
\bigg( \frac{\pi T}{4 m_x} - \frac{11}{12}\bigg( 1-\frac{9m^2_{\rm eq}}{44m^2_x} \bigg)  \bigg).
\ea
For the application in relativistic viscous hydrodynamics, 
it is more useful to use the speed of sound. Applying Eqs.~(\ref{speed-be}) and (\ref{bulk-be1}), one can explicitly show that the ratio is proportional to $(1/3-c_s)^2$, namely
\ba
\label{eq:zeta_be}
{\zeta \over\tau_R}
&\approx& 
T^4 \left({1\over 3} - c_s^2\right)^2 
\left( {2\pi^3 T \over 25 m_{x}} - {4\pi^2 \over 75 } 
\left( 1-\frac{9 m^2_{\rm eq}}{8 m^2_x} \right) \right).\nn\\
&&
\ea

Note the appearance of $T/m_{x}$ in the expression (\ref{eq:zeta_be}).
This is in clear contrast to the Boltzmann statistics case
which does not show such a behavior.
The analysis for the Boltzmann statistics case is identical to the analysis above
except that in place of $J_{n,q}$ we have
\ba
\label{I}
I_{n,q} &=& a_q \int dK  (u \cdot k)^{n-2q} 
(-\Delta_{\mu\nu} k^\mu k^\nu )^q  f_{0,c}(k),\qquad
\ea
where $f_{0,c}(k) = e^{-\beta k^\mu u_\mu}$.
In this case, one gets
\be
\label{zeta-boltzmann}
{\zeta_{\rm Boltz}\over \tau_R}
\approx
T^4 \left({1\over 3} - c_s^2\right)^2
\left(
{60\over \pi^2} - {36m_{x} \over \pi T}
\right).
\ee
The origin of this discrepancy is
the fact that the Bose-Einstein factor behaves like
$f(k)\sim T/E_k$ in the infrared limit, which makes the thermodynamic integral
$J_{-1,0}$ in Eq.~(\ref{bulk-be1}) diverge in the $m_x\to 0$ limit while
$I_{-1,0}$ does not.
As a result, soft
momenta govern the structure of $\zeta/\tau_R$. However, since the
calculation was performed in the relaxation time approximation, which
assumes that $\tau_R$ is independent of energy, 
it may not capture the right
soft physics. A similar behavior was seen 
in Ref.~\cite{Arnold:2006fz}, where QCD bulk viscosity is
studied. The authors claim that the correct behavior of bulk viscosity is
obtained in the relaxation time approximation
by neglecting the infrared divergent term.
But in principle there is no reason why this term
should be ignored within the present framework.

Further, notice that starting from Eq.~(\ref{eq:AW-LHS}), the spin 0 part (the bulk
part) and the spin 2 part (the shear part) of the analysis are totally independent.
Hence, it is possible to generalize the leading order Anderson-Witting equation 
as
\ba
&&\left(k^\mu\partial_\mu 
- {1\over 2}\partial_i m_{\rm eq}^2 
{\partial\over \partial k_i}\right)f_0(x,k) \nn \\
&& \qquad\qquad\qquad
= -{E_k\over \tau_\pi} \Delta f_{\rm s}(x,k)
-{E_k\over \tau_\Pi} \Delta f_{\rm b}(x,k),\qquad
\ea
where $\Delta f_{\rm s}$
and $\Delta f_{\rm b}$ are the shear and bulk parts of $\Delta f$. 
In fact, when the dominant physical
processes for the shear relaxation and the bulk relaxation are different,
this is the most natural form of the Anderson-Witting model.
The analysis of this generalized Anderson-Witting model follows exactly the same
route as for the single $\tau_R$, except that the shear viscosity
and the bulk viscosity have different relaxation times.

As discussed in Refs.~\cite{Jeon:1994if,Jeon:1995zm}, the dominant physical processes for the
shear relaxation and the bulk relaxation can be indeed very different, and the bulk
relaxation can be dominated by the soft sector.
Hence, the appearance of $T/m_x$ is not entirely unnatural given that 
$\tau_\Pi$ can have very different $m_x$ dependence from $\tau_\pi$ and
the bulk relaxation is dominated by the soft number-changing process.

\subsection{Comparison of $\Delta f$ to previous works}

The phase space correction $\Delta f$ in Eq.~(\ref{del-II}) ultimately
comes from solving the first-order Chapman-Enskog approximation.
Hence, it should come as no surprise that Eq.~(\ref{del-II})
is consistent with similar results found in other similar works,
provided that the right expression for the speed of sound is used.
For instance, in Ref.~\cite{Romatschke:2011qp} one finds
that the bulk part of the phase space correction in 
the Boltzmann case is derived to be
\ba
\label{delta-Rom}
\Delta f_R(k) = f_{0,c}(k) \phi_R(k)
\ea
with
\ba
\label{phi-Rom}
\phi_R(k) &=&\tau_R \beta (\partial_i u^i) \bigg((c_{sR}^2-1/3)E_k  \\ \nn
&&-\frac{1}{E_k}\bigg( c_{sR}^2 m_x T \frac{d m_x}{dT} - \frac{m^2_x}{3}\bigg)  \bigg),
\ea
where the speed of sound is $c^2_{sR}=(3+zK_2(z)/K_3(z))^{-1}$, with $z=m_x/T$ and $K_n(z)$ being the modified Bessel functions of the second kind. 
This $\phi_R$ is different than $\phi_b$ in Eq.~(\ref{eq:phi_b_fin})
since $\phi_R$ is a part of $\Delta f$ while $\phi_b$ is a part of
$\delta f$. The phase space correction $\Delta f_R$ is, however,
equivalent to the bulk part of $\Delta f$ in Eq.~(\ref{del-II})
if one uses the speed of sound expression
(\ref{sound}) with $J_{n,q}\to I_{n,q}$.
As mentioned above, this is as it should be since both are solutions of the first-order
Chapman-Enskog approximation.

The big difference between the previous treatments and ours is in computing the bulk viscosity.
The bulk viscosity must be calculated using $\delta f$ and
{\em not} $\Delta f$ as explained in the previous section.
If one uses $\Delta f$ (or $\Delta f_R$) instead of $\delta f$,
the ratio $\zeta/\tau_R$ would be incorrectly calculated.

\section{Transport coefficients in the 14-moment approximation}
\label{sec-evolution}

When a system features a conformal anomaly, first-order 
transport coefficients reveal different
sensitivity to the source of the conformal symmetry violation, as explicitly shown in the previous section. In particular, shear viscosity is fully determined by the dominant energy
scale, which is the temperature $T$, and thus the shear viscosity over its
relaxation time ratio behaves as $T^4$ at leading order in the conformal symmetry breaking, making the effects of scale anomaly negligible. On the other hand, bulk
viscosity over the relaxation time is fully determined by the breaking of conformal symmetry. Such a difference makes it justified to
omit the analysis of shear viscous effects and to evaluate first- and second-order 
transport coefficients related to bulk pressure, because the additional term in Eq.~(\ref{Delta-f1}) 
indeed concerns only the scalar part. The analysis is performed at leading order 
in the conformal breaking parameter
while including the thermal mass consistently.

The bulk pressure is given by Eq.~(\ref{eq:Pi_final}). Noting that Eq.~(\ref{Delta-f1}) can be expressed as
\ba
M \Delta f &=& M \delta f - T^2 \frac{d m^2_{\text{eq}}}{dT^2}
\frac{f_0(1+f_0)}{E_k}  
\frac{\Pi}{\int dK E_k f_0(1+f_0)}, \nn\\
&&
\ea
one can rewrite Eq.~(\ref{eq:Pi_final}) as
\be
\Pi = 
\tilde M \int dK\,\Delta f,
\ee
where
\be
\tilde M =
{MJ_{1,0}\over J_{1,0} - T^2(dm_{\rm eq}^2/dT^2)J_{-1,0}}.
\ee

To obtain the
equation of motion for the bulk pressure, we first take
the time derivative of $\Pi$,
\ba
\label{Pi-eom}
\dot \Pi &=& 
\dot{\tilde{M}} \int dK\,\Delta f \nn\\
&&
+
\tilde{M} \bigg[ \int dK \Delta \dot f 
- \frac{\dot m^2_{\text{eq}}}{2}\int dK \frac{1}{E_k^2} \Delta f\bigg] ,
\ea
where we adopted the notation $\dot A = u^\mu \partial_\mu A$ for an
arbitrary quantity $A$, which reduces to the time derivative
in the rest frame of the fluid.
From the Boltzmann equation 
\be
\left( \tilde k^\mu\partial_\mu - {\cal E}_k \nabla {\cal E}_k\cdot \nabla_k \right)f = 
C[f],
\label{eq:Boltzmann_eq}
\ee
where $C[f]$ is the collision integral, one finds
\ba
\label{delta-eom}
u^\mu \partial_\mu(\Delta f)&=&
\frac{1}{(u\cdot \tilde k)} \bigg[ C[f] - \tilde k^\mu \partial_\mu f_0 
- \tilde k^\mu \nabla_\mu \Delta f \nn\\
&&\qquad
+\frac{1}{2} \nabla \tilde m_x^2 \nabla_k f_0 
+\frac{1}{2} \nabla \tilde m_x^2 \nabla_k \Delta f   \bigg].\qquad
\ea
Inserting the expression (\ref{delta-eom})
to Eq.~(\ref{Pi-eom}) and keeping only leading order terms, that is, terms which are evaluated 
with $\tilde k \to k$, we have
\ba
\label{Pi-eom-1}
\dot \Pi - C &=&
- \tilde{M}\bigg[- \dot\beta \Big(J_{1,0} -  
T^2(dm^2_{\text{eq}}/dT^2) J_{-1,0}\Big) 
\nn\\
&&
+ \frac{\beta}{3} \theta \Big( J_{1,0} - m_x^2 J_{-1,0} \Big)  \bigg] 
+\left( \frac{\dot{\tilde M}}{\tilde M}- \frac{2}{3} \theta \right) \Pi
\nn\\
&&
 -M\Big( \frac{\dot m_{\text{eq}}^2}{2} + \frac{m_x^2}{3} \theta \Big) \rho_{-2} 
- M \rho_{-2}^{\mu\nu} \sigma_{\mu\nu} ,
\ea
where $\theta \equiv \nabla_\mu u^\mu$ and $\sigma_{\mu\nu} = \partial_{\langle\mu} u_{\nu\rangle}$ is the Navier-Stokes shear tensor. In Eq.~(\ref{Pi-eom-1}) we adopted the following notation for the collision term:
\ba
\label{C}
C &=& \tilde M \int dK (u \cdot k)^{-1} C[f]
\ea
and, for the irreducible moments,
\ba
\label{rho-1}
\rho_n =\langle (u^\alpha k_\alpha)^n \rangle_\delta, \qquad
\rho_n^{\mu\nu} = \langle (u^\alpha k_\alpha)^n k^{\langle \mu} k^{\nu \rangle} \rangle_\delta.
\;\;\;
\ea

Evaluating $u_\nu \partial_\mu T^{\mu\nu}=0$ and implementing the formula (\ref{sound}) for the speed of sound squared, one obtains
\ba
\label{beta-dot}
\dot \beta = \frac{ \Pi \theta -\pi^{\mu\nu} \sigma_{\mu\nu} }
{J_{3,0} - T^2 (dm^2_{\text{eq}}/dT^2)J_{1,0}} + c_s^2 \beta \theta .
\ea
Next, calculating time derivatives $\dot{\tilde M}$ and $\dot m^2_{\text{eq}}$, Eq.~(\ref{Pi-eom-1}) simplifies to
\ba
\label{Pi-eom-2}
\dot \Pi - C &=&
- 
\beta\tilde{M} \bigg[\bigg(\frac{1}{3}- c_s^2\bigg) 
\left( J_{1,0} - T^2\frac{dm_{\rm eq}^2}{dT^2}J_{-1,0}\right)\nn\\
&&
+ M J_{-1,0}  \bigg] \theta 
- \Big(\frac{2}{3} + \frac{2c_s^2 aT^2\beta_\lambda}{3\tilde M}-A \Big)\theta \Pi \nn\\
&&
- \pi^{\mu\nu}\sigma_{\mu\nu} A 
+M^2 \rho_{-2}  \theta
- M \rho_{-2}^{\mu\nu} \sigma_{\mu\nu} ,
\ea
where
\ba
\label{A}
A=\tilde M \frac{J_{1,0}- T^2(dm^2_{\text{eq}}/dT^2) J_{-1,0}}{J_{3,0}- T^2(dm^2_{\text{eq}}/dT^2)J_{1,0}} = c_s^2 -\frac{1}{3}\qquad
\ea
with the quantity $(c_s^2-1/3)$ given by Eq.~(\ref{speed-be}).

To close Eq.~(\ref{Pi-eom-2}) in terms of $\Pi$ and $\pi^{\mu\nu}$, one can apply the 14-moment approximation, which allows one to express the irreducible moments by $\Pi$ and $\pi^{\mu\nu}$ as follows:
\ba
\label{mom-1}
\rho_{-2} &=& \gamma^{(0)}_{2} \Pi,\\
\label{mom-2}
\rho^{\mu\nu}_{-2} &=&\gamma^{(2)}_2 \pi^{\mu\nu},
\ea
where the coefficients $\gamma^{(0)}_{2}$ and $\gamma^{(2)}_2$ are combinations of different thermal functions $J_{n,q}$. Their particular forms are presented in Appendix~\ref{moments}. Also, using the Anderson-Witting model for the collision term
\ba
C[f] = - (u\cdot k) \frac{\Delta f}{\tau_R},
\ea
where $\Delta f$ is given by Eq.~(\ref{Delta-f1}), the collision integral becomes
\ba
\label{C1}
C &=& -\frac{\Pi}{\tau_R}.
\ea

Applying the collision term in the relaxation time approximation (\ref{C1}), the irreducible moments, Eqs.~(\ref{mom-1}) and (\ref{mom-2}), and the relation for the speed of sound (\ref{speed-be}) to the evolution equation (\ref{Pi-eom-2}), one obtains
\ba
\label{Pi-eom-3}
\dot \Pi + \frac{\Pi}{\tau_R} &=&
-{\zeta\theta\over\tau_R} - \frac{\delta_{\Pi\Pi}}{\tau_R}\theta \Pi + \frac{\lambda_{\Pi \pi}}{\tau_R} \pi^{\mu\nu}\sigma_{\mu\nu},
\ea
where
\ba
{\zeta\over \tau_R}
&=&
\beta
M^2
\bigg[
{J_{1,0}J_{-1,0}\over J_{1,0} - T^2(dm_{\rm eq}^2/dT^2)J_{-1,0}}\qquad\qquad\nn\\
&&\qquad
-{J_{1,0}J_{1,0}\over J_{3,0} - T^2(dm_{\rm eq}^2/dT^2)J_{1,0}}
\bigg] 
\ea
is identical to the expression obtained in the Chapman-Enskog
approach found in the previous section, Eq.~(\ref{bulk-be1}). The remaining transport coefficients are 
\ba
\label{bulk-del-bose}
\frac{\delta_{\Pi\Pi}}{\tau_R} &=&1 - c_s^2 +M^2 \gamma_2^{(0)} + \frac{2 aT^2\beta_\lambda}{9\tilde M}, \\
\label{bulk-lam-bose}
\frac{\lambda_{\Pi \pi}}{\tau_R} &=&\frac{1}{3} - c_s^2 - M\gamma_2^{(2)}.
\ea
Converting $M$ to the speed of sound and taking $m_0 \to 0$ limit, one gets
\ba
\label{bulk-del-boltz}
\frac{\delta_{\Pi\Pi}}{\tau_R} &\approx& 
\frac{4}{3} \left(1+ \frac{T^2}{2}\frac{dm_{\rm eq}^2}{dT^2}\frac{J_{-1,0}}{J_{1,0}}\right) 
+ \left(\frac{1}{3}-c_s^2 \right) \nn\\
&&
+ \gamma_2^{(0)}  \left({J_{3,0}\over J_{1,0}} - T^2\frac{dm_{\rm eq}^2}{dT^2} \right)^2 
\left( \frac{1}{3}-c_s^2 \right)^2,  \\
\label{bulk-lam-boltz}
\frac{\lambda_{\Pi \pi}}{\tau_R} &\approx&
\left( 1+ \gamma_2^{(2)} \left({ J_{3,0} \over J_{1,0}} - T^2\frac{dm_{\rm eq}^2}{dT^2} \right)  \right) \left( \frac{1}{3}-c_s^2 \right) ,\qquad
\ea
where $\gamma_2^{(0)}$ and $\gamma_2^{(2)}$ are calculated in Appendix \ref{moments} and are given by Eqs.~(\ref{gam-1a-bose}) and (\ref{gam-2a-bose}), respectively. When inserted, one gets the leading orders of the coefficients,
\ba
\label{bulk-del-boltz-1}
\frac{\delta_{\Pi\Pi}}{\tau_R}
&\approx& 
\frac{4}{3}\left(1+ \frac{3}{8\pi}\frac{m_{\rm eq}}{T} 
- \frac{3}{16\pi^2} \frac{m^2_{\rm eq}}{T^2} \right) \nn \\
&&
+ \left(\frac{1}{3}-c_s^2 \right) \left(\frac{6}{15\pi} \frac{T}{m_{\rm eq}} +1 \right) \nn \\
&&
+  0.97 \left( \frac{1}{3}-c_s^2 \right)^2 \frac{T^4}{m_{\rm eq}^4},  \\
\label{bulk-lam-boltz-1}
\frac{\lambda_{\Pi \pi}}{\tau_R} &\approx& 1.05 \left(\frac{1}{3} - c_s^2\right),
\ea
where the numerical factors come from evaluating $g_0 \left(12/15 \right)^2 \approx 0.97$ and $\left(1+12g_2/15 \right) \approx 1.05$ with $g_0$ and $g_2$ given by Eqs.~(\ref{g0}) and (\ref{g2}).
As seen, the coefficient $\delta_{\Pi\Pi}/\tau_R$ is affected by the soft physics even more strongly than the bulk viscosity which is manifested by the factors $1/m_{\rm eq}$ and $1/m_{\rm eq}^4$.

Repeating the same analysis for the Boltzmann gas, which leads simply to replacement of the thermodynamic functions $J_{n,q} \to I_{n,q}$, one obtains the same value of $\zeta_{\rm Boltz}/\tau_R$ as within the Chapman-Enskog approach, Eq.~(\ref{zeta-boltzmann}). The other two coefficients have the forms (\ref{bulk-del-boltz}) and (\ref{bulk-lam-boltz}) with $\gamma_2^{(0)}$ and $\gamma_2^{(2)}$ given by Eqs.~(\ref{gam-1a}) and (\ref{gam-2a}). The explicit expressions in the $m_0 \to 0$ limit are then found to be
\ba
\label{bulk-del-boltz2}
\frac{\delta_{\Pi\Pi,{\rm Boltz}}}{\tau_R} 
&\approx& 
\frac{4}{3}\left( 1+\frac{1}{4} \frac{m^2_{\rm eq,\;c}}{T^2} \right)
+5 \left(\frac{1}{3} - c_s^2\right) \nn\\
&&
-10.8\left(\frac{1}{3} - c_s^2\right)^2 , \\
\label{bulk-lam-boltz2}
\frac{\lambda_{\Pi \pi,{\rm Boltz}}}{\tau_R} &\approx&  1.6 \left(\frac{1}{3} - c_s^2\right) ,
\ea
where the numerical factors were found from $144g_{0c} \approx -10.8$ and $\left(1+12g_{2c} \right) \approx 1.6$ with $g_{0c}$ and $g_{2c}$ written up below Eq.~(\ref{gam-2a}). One can also see from 
Eqs.~(\ref{bulk-del-bose}) and (\ref{bulk-lam-bose}) that when thermal quantities are neglected and the constant mass is kept, we reproduce  $\zeta_{{\rm Boltz}}/\tau_R$, $\lambda_{\Pi \pi,{\rm Boltz}}/\tau_R$ and the two first terms of $\delta_{\Pi \Pi,{\rm Boltz}}/\tau_R$ from Ref.~\cite{Denicol:2014vaa}.

\section{Summary and conclusions}
\label{sec-summary}

In this paper we analyzed the influence of the mean field on fluid
dynamics in weakly interacting systems of a single species, where all
occurring masses are much smaller than the system's temperature. Our main
attention was paid to proper determination of the form of the
nonequilibrium correction to the distribution function which depends 
on the mass varying as the temperature varies. The correction guarantees a consistent
hydrodynamic description which satisfies thermodynamic relations and the conservation of energy
and momentum and furthermore gives an accurate fixing of the temperature through Landau matching.
 The correction
plays a central role in studying thermal dependence of bulk viscous
dynamics. Therefore, we further considered the Anderson-Witting model 
of the Chapman-Enskog approach and 
computed $\zeta/\tau_R$ of single-component Bose-Einstein and Boltzmann
gases. We also derived the evolution equation for the bulk pressure in the
 14-moment approximation and obtained relevant transport coefficients. Both
methods provide the same result for~$\zeta/\tau_R$.

The ratio $\zeta/\tau_R$ obtained for the Boltzmann statistics behaves as
expected, that is, it is given by the nonconformality parameter squared.
When thermal effects are omitted, we reproduce the result from 
Refs.~\cite{Denicol:2014vaa,Florkowski:2015lra}. On the other hand, for very high
temperatures the ratio gets dominated by the $\beta_\lambda$ function. We
also see that in spite of breaking conformal invariance, bulk viscosity
vanishes at some critical temperature where $c_s^2=1/3$. In the case of the Bose-Einstein gas, we
have shown that the leading order term of $\zeta/\tau_R$ is different than expected 
if we neglect either the physical mass or thermal effects. The ratio in this case is strongly redounded by the infrared
physics, which introduces an additional energy-scale-dependent factor $T/m_x$. We suspect that the relaxation time approximation
used here does not include the entire microscopic physics of a quantum gas, in
particular, it is insensitive to phenomena at the soft scale.
Therefore, we conclude that to compute the bulk viscosity over its
relaxation time for quantum gases of Bose-Einstein statistics, one needs to
use more advanced methods and solve an integral equation. It can be done
starting from either the linearized Boltzmann equation or Kubo formulas, in
which case note that the formula for the bulk relaxation time was recently found
\cite{Czajka:2017bod}.

\section*{Acknowledgments}
It is a pleasure to thank G. Denicol, J. Kapusta, and J.-F. Paquet for useful discussions.  This work is supported in part by the Natural Sciences and Engineering Research Council of Canada, and by the US DOE, under Contract No. DE-SC0012704. In addition, we gratefully acknowledge support from the program Mobility Plus of the
Polish Ministry of Science and Higher Education (A. C. ),  from the Canada Council for the Arts through its Killam Research Fellowship program (C. G.), and from the Goldhaber Distinguished Fellowship program from Brookhaven Science Associates (C. S.).

\appendix

\section{Components of the energy-momentum tensor correction}
\label{components}

The correction to the energy-momentum tensor is
\ba
\label{visc-delta1}
\Delta T^{\mu\nu} = \int dK k^\mu k^\nu  \delta f + \frac{\partial
T_0^{\mu\nu}}{\partial m^2_{\text{eq}}} \Delta m^2_{\text{th}}
\ea
and its particular components are derived as follows. For $\mu=\nu=0$, one
gets
\ba
\label{T-00a_not_used}
\Delta T^{00} &=& \int dK E_k^2 \delta f + \frac{\Delta m^2_{\text{th}}}{2}
\int dK  f_0 \nn\\
&&
- \frac{\Delta m^2_{\text{th}}}{2} \beta\int dK E_k  f_0(1+f_0) -\Delta U
\nn\\
&=& \int dK E_k^2 \Delta f,
\ea
where the condition on $\Delta U$, given by (\ref{U1}), and Eq.~(\ref{Delta-f}) have been used. 
 Using Eq.~(\ref{Delta-f1}) for $\Delta f$, one has
 \ba
 \label{T-00a}
 \Delta T^{00}
 &=& \int dK E_k^2 \Delta f,
 \nonumber\\
 &=&
 \int dK E_k^2 
 \bigg[ \delta f \nn \\
&&
- T^2 \frac{d m^2_{\text{eq}}}{dT^2}
 \frac{f_0(1+f_0)}{E_k}  
 \frac{\int dK \delta f}{\int dK E_k f_0(1+f_0)}.
 \bigg]
 \nn\\
 &=&
 \int dK 
 \left( E_k^2 - T^2 \frac{d m^2_{\text{eq}}}{dT^2} \right)
 \delta f
 \ea

Analogously, one gets the momentum density
variation
\ba
\label{T-0ia}
\Delta T^{0i} &=& \int dK E_k k^i \delta f 
- \frac{\Delta m^2_{\text{th}}}{2} \beta\int dK k^i  f_0(1+f_0) \nn\\
&=& \int dK E_k k^i \Delta f .
\ea
The stress tensor variation is
\ba
\label{T-ija}
\Delta T^{ij} &=& \int dK k^i k^j\delta f 
- \frac{\Delta m^2_{\text{th}}}{2} \beta\int dK \frac{k^i k^j}{E_k}
f_0(1+f_0) \nn\\
&&
-\frac{\Delta m^2_{\text{th}}}{2} \int dK \frac{k^i k^j}{E_k^2} f_0 
+\delta^{ij} \Delta U\;\; \nn \\
&=& 
\int dK k^i k^j \Delta f 
-\frac{\Delta m^2_{\text{th}}}{2} \int dK \frac{k^i k^j}{E_k^2} f_0  \nn\\
&&
+\delta^{ij} \frac{\Delta m^2_{\text{th}}}{2}
\int dK f_0,
\ea
where Eq.~(\ref{Delta-f1}) has been applied. Equations~(\ref{T-00a}) -
(\ref{T-ija}) correspond to Eqs.~(\ref{T-00}) - (\ref{T-ij}).
Among all these expressions, $\Delta T^{ij}$ needs further simplifications
to show how one can obtain Eq.~(\ref{T-ij-del}). The second and the third
terms of the first line in Eq.~(\ref{T-ija}) may be combined to get
\ba
\label{T-ijaa}
\Delta T^{ij} &=& \int dK k^i k^j\delta f 
- \frac{\Delta m^2_{\text{th}}}{2} \int dK k^i k^j \partial_{E_k}
\bigg(\frac{f_0}{E_k}\bigg) \nn\\
&&+\delta^{ij} \Delta U.
\ea
Next, using $\partial_{E_k}(\dots) =
\frac{E_k}{k}\partial_k(\dots)$ and then integrating by parts leads to
\ba
\label{T-ijaaa}
\Delta T^{ij} &=& \int dK k^i k^j\delta f 
- \delta^{ij} \frac{\Delta m^2_{\text{th}}}{2} \int dK f_0
+\delta^{ij} \Delta U \nn\\
& =& \int dK k^i k^j\delta f,
\ea
where the condition (\ref{U1}) has been used.

\section{Details of the thermodynamic integrals}
\label{bessel-fun}

\subsection{Boltzmann statistics}
\label{boltz-details}

Our strategy to evaluate the integrals with the Boltzmann statistics is to use the integral representation of
the modified Bessel functions of the second kind
\ba
\label{bessel1}
K_n(z) = 
\int_0^\infty d\theta \cosh (n\theta) \exp{(-z \cosh \theta)},
\ea
We will also need the Bickley functions defined by
\ba
\label{bickley}
\text{Ki}_r(z) = \int_0^\infty d\theta \frac{\exp{(-z \cosh
\theta)}}{(\cosh \theta)^r}.
\ea
We will need the following series in the small $z$ limit
\ba
\label{k1}
K_1(z) &\approx& \frac{1}{z} -\frac{z}{4} \big(1-2\gamma_E + \ln 4 - 2\ln z  \big),
\\
\label{k2}
K_2(z) &\approx& \frac{2}{z^2}-\frac{1}{2} +\frac{z^2}{32} \big(3-4\gamma_E + 2\ln
4 - 4\ln z  \big), \qquad\\
\label{k3}
K_3(z) &\approx& \frac{8}{z^3}-\frac{1}{z} +\frac{z}{8}, \\
\label{k4}
K_4(z) &\approx& \frac{48}{z^4} - \frac{4}{z^2} +\frac{1}{4},\\
\label{k5}
K_5(z) &\approx& \frac{384}{z^5}-\frac{24}{z^3} +\frac{1}{z}, 
\ea
where $\gamma_E=0.577$ is the Euler constant
and $\ln 4=1.386$ and the higher order terms in $z$
were neglected. 
For the Bickley function \cite{Blair}, we need
\ba
{\rm Ki}_1(z) 
&\approx&
{\pi\over 2} -z(1 -\gamma_E - \ln(z/2))
\ea

Using $|{\bf k}| \equiv k = m_x\sinh\theta$,
the thermodynamic functions $I_{n,q}$, defined by (\ref{I}) and evaluated in the fluid rest frame,
can be expressed as
\ba
&&I_{n,q} (T,z) = a_q  \frac{T^{n+2} z^{n+2}}{2\pi^2} \nn\\
&& \times
\int_0^\infty d\theta (\cosh \theta)^{n-2q} (\sinh \theta)^{2q+2} \exp{(-z
\cosh \theta)},\qquad
\ea
where $z=\frac{m_x}{T}$ and $a_q=1/((1+2q)!!)$.
Using $\cosh x = (e^{x} + e^{-x})/2$ and $\sinh x = (e^{x}-e^{-x})/2$, 
and the definition~(\ref{bessel1}),
these integrals can be expressed in terms of modified Bessel functions of the second kind.

Let us consider $I_{3,0}$ first. After the angle integral, we have
\ba
I_{3,0}
&=&
{1\over 2\pi^2} \int_0^\infty dk\, k^2\, E_k^2\, e^{-E_k/T}.
\ea
Using $k = m_x \sinh\theta$, this becomes
\ba
I_{3,0}
&=&
{m_x^5\over 2\pi^2} \int_0^\infty d\theta\, \sinh^2\theta \cosh^3\theta \, e^{-z\cosh\theta},
\ea
where $z = m_x/T$. By using $\cosh x = (e^{x} + e^{-x})/2$ and $\sinh x = (e^{x}-e^{-x})/2$ 
and the definition~Eq.~(\ref{bessel1}), one gets
\ba
I_{3,0}
&=&
-{m_x^5\over 32\pi^2} \left(2 K_1(z)-K_3(z)-K_5(z)\right)\nn\\
&\approx &
{m_x^5\over 32\pi^2} \left(
\frac{384}{z^5}-\frac{16}{z^3} \right) 
= \frac{12T^5}{\pi^2} \left( 1-\frac{z^2}{24} \right). \qquad
\ea

The other useful integrals are found in a similar way:
\ba
I_{1,0} &\approx&  \frac{T^3}{\pi^2}  \left( 1 - \frac{z^2}{4}  \right),
\\
I_{-1,0} &\approx& \frac{T}{2\pi^2} \left(1 - \frac{z\pi}{2}\right),
\\
I_{3,1} &\approx& \frac{4T^5}{\pi^2} \left( 1 - \frac{z^2}{8}  \right),
\\
\label{I00}
I_{0,0} &\approx& \frac{T^2}{2\pi^2} \left( 1 -\frac{z^2}{4} \big(1-2\gamma + \ln 4 - 2\ln z  \big) \right),\;\;\;\;
\ea
where $I_{0,0} \equiv q_0$ is needed for the thermal mass evaluation.
For $\epsilon + P$, we have
\ba
\epsilon + P &\approx& {T^4\over \pi^2} \left(4 - {z^2\over 2} + {z^4\over 16}\right).
\ea

\subsection{Bose-Einstein statistics}
\label{bose-details}

The thermodynamic integrals for the Bose-Einstein gas are defined by Eqs.~(\ref{J}): 
\ba
J_{n,q} &=& a_q\int dK  (u^\mu k_\mu)^{n-2q} 
(-\Delta_{\mu\nu} k^\mu k^\nu )^q [ f_0(1+f_0)].\nn\\
&&
\ea
In the fluid rest frame and after the angle integrals, $J_{n,q}$ becomes
\ba
J_{n,q} &=&
{a_q\over 2\pi^2 } \int_0^\infty dk\, {k^2\over E_k}\, F_{n,q}(E_k)\, f_0(E_k)(1+f_0(E_k))\nn\\
&&
\ea
and 
\ba
F_{n,q}(E_k) = E_k^{n-2q}k^{2q}
= E_k^{n-2q}(E_k^2 - m_x^2)^{q}.\qquad
\ea
Using $\partial_k f_0(E_k) = -{k\over TE_k}f_0(E_k)(1+f_0(E_k))$ and integrating by parts,
we can rewrite the above as
\ba
J_{n,q} 
&=&
{ a_qT\over 2\pi^2} \int_0^\infty dk\, f_0(E_k) \, \partial_k \left(k F_{n,q}(E_k)\right).\qquad
\ea
Changing the integration variable to $E_k$, we further get
\ba
J_{n,q} 
&=&
{ a_qT\over 2\pi^2} \int_{m_x}^\infty dE_k\, \, G_{n,q}(E_k) \, f_0(E_k),
\label{eq:IntGnq}
\ea
where 
\ba
&&G_{n,q}(E_k) =
(E_k/k)\partial_k(k F_{n,q}(E_k))
\nn \\
&&
\nn \\
&&
={(E_k^2-m_x^2)^q E_k^{n-2q-1}((n+1)E_k^2 + m_x^2(2q-n))\over \sqrt{E_k^2-m_x^2}}
\qquad \;\;
\ea
using $k = \sqrt{E_k^2-m_x^2}$.

Our strategy to evaluate this integral is to separate the high momentum
contribution and the low momentum contribution.
We know how to evaluate
\ba
\int_{m_x}^\infty dE_k E_k^l f_0(E_k)
&=&
T^{l+1} \int_z^\infty dx\, x^l \sum_{n=1}^\infty e^{-nx}\qquad\qquad
\ea
where $x = E_k/T$ and $z = m_x/T$
in terms of the polylogarithmic functions ${\rm Li}_{n}(z)$.
Hence, we first expand the square root in \(m_x^2/E_k^2\) and identify
the non-negative power terms in \(E_k\). Denoting 
the collection of such terms as $H_{n,q}(E_k)$, we then separate the integral as
\ba
J_{n,q} &=&
{a_qT\over 2\pi^2 } \int_{m_x}^\infty dE_k\, H_{n,q}(E_k)\, f_0(E_k)
\nn\\
&&+
{a_qT\over 2\pi^2 } \int_{m_x}^\infty dE_k\, (G_{n,q}(E_k) - H_{n,q}(E_k))\, f_0(E_k)\nn\\
&&
\ea
One can show that the reminder
$G_{n,q}(E_k)-H_{n,q}(E_k) = O(1/E_k^3)$ for all $n$ and $q$.
Then expanding $f_0$ in the small $E_k/T$ limit,
\ba
f_0(E_k)
&=&
\frac{T}{E_k}-\frac{1}{2}+\frac{E_k}{12T}+O\left((E_k/T)^3\right)
\ea
we can keep the first three terms in the integrand to calculate the soft
contribution.
This integral can usually be exactly evaluated.

Let us consider $J_{3,0}$. From Eq.~(\ref{eq:IntGnq}), we have
\ba
J_{3,0}
&=&
{T\over 2\pi^2} 
\int_{m_x}^\infty dE_k\, 
{
4 E_k^4 - 3m_x^2 E_k^2
\over \sqrt{E_k^2 - m_x^2}}
f_0(E_k).
\ea
Expanding the square-root in powers of $m^2_x/E_k^2$, we get 
\be
G_{3,0}(E_k) = 4 E_k^3 - m_x^2 E_k + O(1/E_k^3).
\ee
We can then separate the hard and the soft parts
\ba
&&J_{3,0}
=
{T\over 2\pi^2} 
\int_{m_x}^\infty dE_k\, 
\left(
4 E_k^3 - m_x^2 E_k
\right)
f_0(E_k)
\nn\\
&&{}
+
{T\over 2\pi^2} 
\int_m^\infty dE_k
\left[
{4 E_k^4 - 3m_x^2 E_k^2 \over \sqrt{E_k^2 - m_x^2}}
-4 E_k^3 + m_x^2 E_k
\right]
f_0(E_k).\nn\\
&&
\ea
Since the square bracket behaves like $1/E_k^3$, we can use
$f_0(E_k) \approx T/E_k - 1/2 + E_k/12T$ to evaluate the second
integral. It is
\be
J_{n,q}^{\rm soft}
\approx
{T^5\over 2\pi^2} 
\left(
{z^3\over 3} - {3z^4\over 16} + {7z^5\over 180}
\right)
\ee
with $z = m_x/T$.
The hard part is
\ba
J^{\rm hard}_{3,0}
&=&
{T\over 2\pi^2} 
\int_{m_x}^\infty dE_k\, 
\left(
4 E_k^3 - m_x^2 E_k
\right)
f_0(E_k)
\nn\\
&=&
{T^5\over 2\pi^2} 
\int_z^\infty dx\, 
\left(
4 x^3 - z^2 x
\right)
{1\over e^{x}-1}
\nn\\
&=&
{T^5\over 2\pi^2} 
\Big(
24\,{\rm Li}_4(e^{-z})
+
24z\,{\rm Li}_3(e^{-z})
+
11z^2\,{\rm Li}_2(e^{-z})\nn\\
&&
+
3z^3\,{\rm Li}_1(e^{-z})
\Big)
\nn\\
&=&
{T^5\over 2\pi^2} 
\left(
\frac{4 \pi ^4}{15}-\frac{\pi ^2 z^2}{6}-\frac{z^3}{3} 
+ {z^4\over 4} 
- {7z^5\over 180}
O\left(z^6\right)
\right).\nn\\
&&
\ea
Adding the two yields
\ba
J_{3,0} &=& J_{3,0}^{\rm hard} + J_{3,0}^{\rm soft} \approx
{T^5\over 2\pi^2} \left( \frac{4 \pi ^4}{15}-\frac{\pi ^2 z^2}{6}+\frac{z^4}{16} \right).\qquad\;\;
\label{eq:J30_eval}
\ea
This formula works better than 1 part in $10^4$ up to $z = m_x/T = 1$.

The usual way of evaluating Bose-Einstein integrals is to use 
modified Bessel functions of the second kind:
\ba
J_{3,0} &=&
{T\over 2\pi^2}
\int_0^\infty dk {k\over E_k} {4E_k^4 - 3m_x^2 E_k^2\over k} \sum_{n=1}^\infty e^{-nE_k/T}\nn\\
&=&
{T m_x^4\over 2\pi^2}
\sum_{n=1}^\infty
\left(
{1\over 2}K_4(nz)
+ {1\over 2}K_2(nz)
\right). \qquad
\ea
Using the small $x$ expansion of $K_n(x)$ and collecting only the terms
converging under the infinite sum, we get
\ba
J^{\rm Bessel}_{3,0}
&\approx &{T m_x^4\over 2\pi^2}
\sum_{n=1}^\infty\left({24\over n^4 z^4} - {1\over n^2 z^2}\right)\nn\\
&=&
{T^5\over 2\pi^2}
\left(\frac{4 \pi^4}{15}-\frac{\pi^2 z^2}{6}\right).
\label{eq:J30_bessel_eval}
\ea
which gets only the first two terms.

The useful integrals are then found using the former method:
\ba
J_{1,0}&\approx&{T^3\over 6}\left( 1 -\frac{3  z}{2\pi} +\frac{3z^2}{4\pi^2} \right), 
\\
J_{3,1} &\approx& {2T^5\pi^2\over 45}\left(1-\frac{15 z^2}{8\pi^2}\right),
\\
J_{-1,0} &\approx&{1\over 8\pi} \frac{T}{z} \left(1-\frac{2z}{\pi} + \frac{z^2}{6} \right).
\ea
These formulas provide very good approximation up to $z = 1$.
In evaluating $J_{-1,0}$ one would expect to use the Bickley functions equivalently but this method does not work because the sum cannot be easily evaluated, even for the leading behavior.

For the enthalpy, one gets
\ba
\epsilon + P &\approx& { 2\pi^2 T^4\over 45}\left( 1-\frac{15 z^2}{8 \pi ^2}\right)
\ea
and for the thermal mass
\ba
q_0 &\approx&{T^2\over 12}\left(1 - {3 z\over \pi}\right).
\ea

\section{Irreducible moments}
\label{moments}

To express the irreducible moments of the distribution function one can apply the Grad's 14-moment approximation, where the correction to the distribution function of the Bose-Einstein gas is a generalization of the Boltzmann one, shown in  \cite{Israel:1979wp,Denicol:2014vaa,Florkowski:2015lra}, and takes the form
\ba
\label{mom-ff}
\delta f &=& f_0(1+f_0) 
\Big[ E_0 + B_0 m_x^2 +D_0 (u\cdot k) - 4B_0 (u \cdot k)^2 \Big] \Pi \nn\\
&&+f_0(1+f_0)  B_2 p^\alpha p^\beta \pi_{\alpha\beta} .
\ea
The coefficients $E_0$, $B_0$, $D_0$, and $B_2$ are functions of $m_x$, $T$, and $u^\mu k_\mu$ and they read
\ba
\label{B2}
B_2 &=& \frac{1}{2J_{4,2}},\\
\frac{D_0}{3B_0} &=& -4 \frac{J_{3,1}J_{2,0}-J_{4,1}J_{1,0}}{J_{3,0}J_{1,0}-J_{2,0}J_{2,0}} 
\equiv- C_2,\\
\frac{E_0}{3B_0} &=& 
m_x^2 + 4 \frac{J_{3,1}J_{3,0}-J_{4,1}J_{2,0}}{J_{3,0}J_{1,0}-J_{2,0}J_{2,0}} 
\equiv -C_1,\\
B_0 &=& - \frac{1}{3C_1 J_{2,1}+ 3C_2 J_{3,1}+3J_{4,1}+5J_{4,2}} ,
\ea
where terms related to the particle diffusion have been dropped. Therefore, the irreducible moments $\rho_{-n}$ and $\rho^{\mu\nu}_{-n}$ can be expressed by $\Pi$ and $\pi^{\mu\nu}$ 
as follows:
\ba
\rho_{-n} &=& \gamma^{(0)}_{n} \Pi,\\
\rho^{\mu\nu}_{-n} &=& \gamma^{(2)}_n \pi^{\mu\nu},
\ea
where the coefficients $\gamma^{(0)}_{n}$ and $\gamma^{(2)}_n$ are
\ba
\label{gam-1}
\gamma^{(0)}_{n} &=& (E_0 + B_0 m_x^2) J_{-n,0} + D_0 J_{1-n,0} -4B_0 J_{2-n,0},\qquad\\
\label{gam-2}
\gamma^{(2)}_n &=& \frac{J_{4-n,2}}{J_{4,2}}.
\ea
Only $\gamma^{(0)}_{2}$ and $\gamma^{(2)}_{2}$ are needed here.

Using the prescription shown in Appendix \ref{bessel-fun} for evaluating relevant thermodynamics functions, one finds the leading order terms of coefficients $E_0$, $B_0$, and $D_0$, which are
\ba
E_0 \approx \frac{e_0}{z^2T^4} , \qquad
D_0 \approx \frac{d_0}{z^2T^5},\qquad
B_0 \approx \frac{b_0}{ z^2 T^6},  \qquad
\ea
where
\ba
e_0 &=& \frac{48\pi^2(\pi^8-10125 \zeta(3)\zeta(5))}{5(19\pi^6 \zeta(3)-2835\zeta^3(3)-300\pi^4\zeta(5))} \nn\\
&\approx& 22.36, \\
d_0 &=& \frac{-216\pi^4(\pi^2\zeta(3)-25\zeta(5))}{19\pi^6 \zeta(3)-2835\zeta^3(3)-300\pi^4\zeta(5)}\nn\\
&\approx& -22.29, \\
b_0 &=& \frac{3\pi^2(\pi^6-405\zeta^2(3))}{19\pi^6 \zeta(3) -2835 \zeta^3(3)-300\pi^4 \zeta(5)}\nn\\ 
&\approx& -0.84.
\ea

Therefore, the leading orders of $\gamma^{(0)}_{2}$ and $\gamma^{(2)}_{2}$ are
\ba
\label{gam-1a-bose}
\gamma^{(0)}_{2} &\approx& \frac{g_0}{z^4T^4} ,\\
\label{gam-2a-bose}
\gamma^{(2)}_2 &\approx& \frac{g_2}{T^2} 
\ea
where 
\ba
\label{g0}
g_0 &=& \frac{32(\pi^8-10125 \zeta(3)\zeta(5))}
{5(19\pi^6 \zeta(3)-2835\zeta^3(3)-300\pi^4\zeta(5))} \nn\\
&\approx&  1.51,\\
\label{g2}
g_2&=& \frac{\zeta(3)}{20\zeta(5)} \approx 0.06
\ea

For the Boltzmann statistics, one needs to change in all integrals $J_{n,q} \to I_{nq}$ and $M \to M_c$. The leading order results for the classical gas are 
\ba
E_0 \approx \frac{\pi^2}{2T^4} , \qquad
D_0 \approx -\frac{\pi^2}{3T^5},\qquad
B_0 \approx -\frac{\pi^2}{96 T^6},\qquad
\ea
which lead to
\ba
\label{gam-1a}
\gamma^{(0)}_{2} &\approx& \frac{g_{0c}}{T^4},\\
\label{gam-2a}
\gamma^{(2)}_2 &\approx& \frac{g_{2c}}{T^2},
\ea
where $g_{0c} =-(5+12\gamma_{E}-12\ln 2) \approx -0.075 $ and $g_{2c} = 1/20$.

\end{document}